\documentclass[preprint,12pt]{elsarticle}
\usepackage{graphicx}
\usepackage{dcolumn}
\usepackage{bm}
\usepackage{float}
\usepackage{amsmath,amssymb,amsfonts}
\usepackage{epstopdf}
\usepackage{diagbox}
\usepackage{enumitem}
\usepackage{multirow}
\usepackage{hyperref}
\usepackage{color}  
\usepackage{todonotes}
\usepackage{comment}

\makeatletter
\def\ps@pprintTitle{%
	\let\@oddhead\@empty
	\let\@evenhead\@empty
	\let\@oddfoot\@empty
	\let\@evenfoot\@oddfoot
}
\makeatother

\usepackage{overpic}
\definecolor{darkGreen}{rgb}{0,0.45,0}
\definecolor{darkBlue}{rgb}{0,0,0.7}
\definecolor{darkRed}{rgb}{0.76, 0.13, 0.28}
\hypersetup{
    colorlinks=true, 
    linktoc=all,    
    linkcolor=darkBlue, 	
    citecolor=darkGreen,
    urlcolor = darkRed,
}

\renewcommand{\d}{{\mathrm{d}}}

\usepackage[margin=2.3cm]{geometry}
\begin{document}
\begin{frontmatter}

\title{Order and Chaos in Systems of Coaxial Vortex Pairs}

\author[cm]{Christiana Mavroyiakoumou\corref{cor1}}%
\ead{cm4291@cims.nyu.edu}
\affiliation[cm]{organization={Courant Institute, Applied Math Lab, New York University},
city={New York}, 
state={NY},
postcode={10012}, 
country={USA}}
\cortext[cor1]{Corresponding author}

\author[ws]{Wenzheng Shi}%
\ead{wenzhengshi@nyu.edu}
\affiliation[ws]{organization={Courant Institute, New York University},
city={New York}, 
state={NY},
postcode={10012}, 
country={USA}}

\date{\today}

\begin{abstract}
Systems of coaxial vortex pairs in an inviscid flow give rise to complex dynamics, with motions ranging from ordered to chaotic. This complexity arises due to the problem's high nonlinearity and numerous degrees of freedom. We analyze the periodic interactions of two vortex pairs with the same absolute strength moving along the same axis and in the same direction. We derive an explicit formula for the leapfrogging period, considering different initial sizes and horizontal separations, and find excellent quantitative agreement with the numerically computed leapfrogging period. We then extend our study to three coaxial vortex pairs with differing strengths, exploring a broad range of initial geometric configurations, and identify conditions that lead to escape to infinity, periodic or quasi-periodic leapfrogging, and chaotic interactions. We also quantify the occurrence of periodic leapfrogging, revealing that the system transitions to two subsystems when vortex pairs have dissimilar strengths and sizes. By performing a sensitivity analysis using neural networks, we find that the initial horizontal separation between the vortex pairs has the most significant effect on the leapfrogging period.

\end{abstract}

\end{frontmatter}

\section{Introduction} \label{sec:intro}


Vortex rings, as coherent structures, are frequently studied in fluid dynamics and appear in various application fields. A variety of vortex ring configurations~\cite{shariff1992vortex} have been investigated using analytical \cite{borisov2013dynamics, blackmore2000transition}, numerical \cite{konstantinov1994chaotic, konstantinov1997numerical}, and experimental \cite{martynenko1989question} methods. 
The primary focus has been on coaxial vortex rings, in which all vortex rings share a common central axis of symmetry and move in the same direction.
Sets of coaxial vortex rings are prevalent in many physical and engineered fluid systems. Notably, in nature, such interactions are observed in the wakes of hovering birds and insects, where wing strokes generate chains of coaxial vortex rings~\cite{rayner1979vortex, ellington1984aerodynamics,heinzel1987travelling}. 
Similarly, aquatic animals such as squids and jellyfish, which utilize jet propulsion mechanisms, leave behind a trail of successive vortex rings as they propel themselves through water~\cite{linden2004optimal,tallapragada2013dynamics,siekmann1963pulsating,gordon2017animal}. 
The hydrodynamic properties that lead to this jet-propulsive mechanism have been studied theoretically in Siekmann  \cite{siekmann1963pulsating}. There, a formula for the thrust produced by an axisymmetric body whose wake takes the form of a set of coaxial vortex rings is developed. 
In Tallapragada and Kelly \cite{tallapragada2013dynamics}, they model the motion of multiple coaxial vortex rings and their influence on the movement of axisymmetric bodies. This mimics the locomotion of certain jellyfish in a fluid. 
Vortex-producing pulsed jets have drawn interest as a means
of propulsion for small underwater vehicles~\cite{mohseni2006pulsatile}.
Understanding the principles of coaxial vortex ring interactions can help scientists and engineers design biomimetic devices that resemble the propulsion and maneuvering of animals \cite{triantafyllou2004review}.

In real flows, the filament cores of vortices have finite sizes, and the effects of viscosity and three-dimensionality induce diffusion of vorticity within the flow. At high Reynolds numbers, a two-dimensional model of $N$ line vortices provides an idealized dynamical system, which can offer insights into the more generalized three-dimensional vortex ring problem. This model is commonly used \cite{helmholtz1858integrale,newton2002n,aref2007point,aref2010150}, but an alternative Hamiltonian approach also describes these interactions \cite{kirchhoff1883vorlesungen,behring2019stability}; in this work, we use both methods. The dynamics of $N = 1, 2, 3$ vortices are integrable, leading to regular motions, while the dynamics of four vortices, without imposed symmetry, are typically chaotic \cite{newton2002n}, exhibiting the classic sensitivity to initial conditions. The particular case considered in this work—multiple pairs of vortices of equal and opposite strength, moving along the same axis—imposes symmetry with zero total circulation. For $N = 4$, with zero total circulation and zero impulse, the system is known to be integrable. However, for $N > 4$ and systems involving three or more coaxial vortex pairs, the system becomes non-integrable even with this symmetry and is expected to exhibit chaotic dynamics.

Nevertheless, there are still cases where the system does not exhibit chaos, showing instead regular motion.
An example of a regular motion is the so-called
\textit{leapfrogging} phenomenon~\cite{love1893motion}. In this leapfrogging process, the trailing vortex ring shrinks and accelerates while the leading ring widens and decelerates. The trailing ring then passes through the leading one, and the leapfrogging process repeats as both vortex rings move forward. In the case of two vortex rings this is a common phenomenon but it is rarer in the case of three coaxial vortex rings, occurring only under certain conditions, which depend on the strengths, widths, and initial horizontal separations between the rings. The transition from ordered (periodic or quasi-periodic leapfrogging with multiple harmonics) to chaotic motions depends on the interplay of these various parameters~\cite{borisov2013dynamics,konstantinov1997numerical,meleshko1994ordered,khanin1982quasi}. 

In Blackmore and Knio~\cite{blackmore2000transition}, Kolmogorov--Arnold--Moser (KAM) theory and Poincar\'e-Birkhoff theory are used to prove that when the distances between vortex rings are sufficiently small relative to their mean radius, certain initial configurations can lead to periodic or quasi-periodic motion. 
As the separation between the vortex rings increases relative to their size, the system transitions to chaotic behavior.
In Konstantinov~\cite{konstantinov1994chaotic} the interaction of up to five vortex rings, initially placed at the same distance, $\rho_0$, from the system's center is investigated.
The parameter $\rho_0$ controls both the initial separations of the rings and their relative sizes, serving as the main control parameter \cite{konstantinov1994chaotic, konstantinov1997numerical}.
For three vortex rings, at small $\rho_0$, the rings exhibit quasi-periodic motion. A slight increase in $\rho_0$ induces chaotic interactions, which subsequently transition back to ordered motions as $\rho_0$ increases further.
In general, the system alternates between ordered and chaotic regimes for different $\rho_0$ intervals. 
In these studies, the vortex rings are assumed to have identical strengths.

In the current work, we fix the strengths of the first two vortex pairs while varying the strength of the third. Additionally, we explore a broader range of initial geometric configurations, allowing the third vortex pair to vary in size and initial distance from the other two pairs. This setup enables a more detailed investigation of the transition between regular and chaotic dynamics in multi-pair vortex systems.
For the vortex pairs that undergo periodic leapfrogging motions, we are interested in obtaining the period of leapfrogging as a function of the system's parameters and initial configurations. The three-vortex-pair case does not allow for an analytical derivation of the leapfrogging period; at least not with the same theoretical tools employed for the two-vortex-pair case. However, by leveraging neural networks, we perform a sensitivity analysis that allows us to extract the main parameters that determine the period of leapfrogging motions.

The structure of the paper is as follows. Section~\ref{sec:model} describes our model in terms of point vortices, similar to the one in Mavroyiakoumou and Berkshire~\cite{mavroyiakoumou2020collinear} but generalized to the case of $N_{\mathrm{p}}$ vortex pairs. Section~\ref{sec:twoRings} focuses on the interaction between two vortex pairs, revisiting the criterion for periodic leapfrogging motions and determining the period of the leapfrogging motion based on the initial sizes and horizontal separations of the pairs. In Section~\ref{sec:threeRings}, we extend the investigation to three interacting vortex pairs and analyze the resulting dynamics given different parameters and initial configurations, and in Section~\ref{sec:sensitivity} we perform a sensitivity analysis on the leapfrogging period based on a six-dimensional parameter space. Section~\ref{sec:conclusions} gives the conclusions.

\section{Equations of motion}\label{sec:model}

\begin{figure}[htpb]
 \centering
 \includegraphics[width=.7\textwidth]{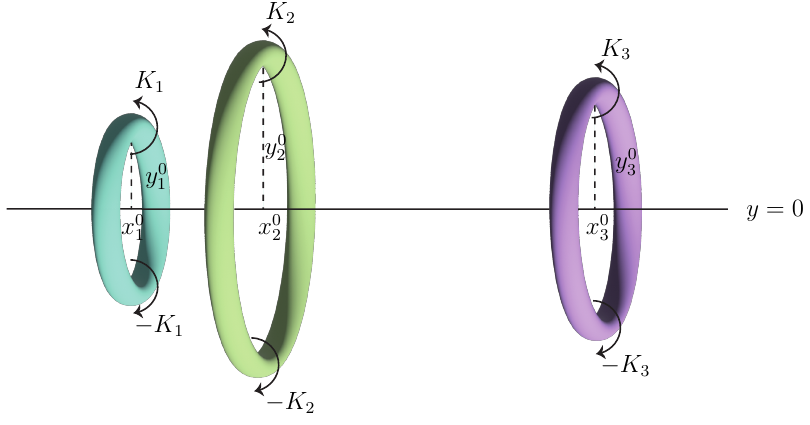}
      \caption{Geometry of three vortex rings of different vortex strengths $K_1$, $K_2$, $K_3$, respectively. The rings with initial radius $y_j^0$ are located at $x_j^0$ along the $x$-axis, with $j=1,2,3$. In this work, we model each vortex ring as a vortex pair of counter-rotating point vortices.}
    \label{fig:schematic}
\end{figure}

We model each vortex ring $j$ as a pair of counter-rotating point vortices of equal strength $K_j$ but with opposite sign, located at $(x_j,y_j)$ and $(x_j,-y_j)$. 
The point vortex at $y_j>0$ has a counterclockwise circulation ($K_j>0$) and the point vortex at $y_j<0$ has a clockwise circulation ($-K_j<0$) associated with it. This is shown schematically in Fig.~\ref{fig:schematic} for three vortex rings, each with a different vortex strength $K_1$, $K_2$, and $K_3$. When $K_i$ and $K_j$ have the same sign, they travel in the same direction, 
and each vortex induces a transverse motion component in each of the other vortices.
The point-vortex model is a widely used simplified method for a two-dimensional incompressible inviscid fluid where the vorticity is confined to a discrete set of moving points~\cite{helmholtz1858integrale}. We can thus write a system of ordinary differential equations that describe the locations of the point vortices evolving over time.

We consider $N_\mathrm{p}$ interacting vortex pairs, which corresponds to  $N=2N_{\mathrm{p}}$ point vortices. The horizontal and vertical components of the induced velocity on vortex $j$ from the remaining $N-1$ point vortices are given by~\cite{mavroyiakoumou2020collinear}:
\begin{align}
     \dfrac{\mathrm{d}x_j}{\mathrm{d}t}&=\frac{K_j}{4\pi y_j} +\sum\limits_{\substack{m=1 \\ m\neq j}}^{N}\frac{K_m}{2\pi}\left[\frac{y_m-y_j}{(x_m-x_j)^2+(y_m-y_j)^2} +\frac{y_m+y_j}{(x_m-x_j)^2+(y_m+y_j)^2}\right],\label{eq:dxdt}\\
 \dfrac{\mathrm{d}y_j}{\mathrm{d}t}&=\hspace{1.25cm}\sum\limits_{\substack{m=1 \\ m\neq j}}^{N}\frac{K_m}{2\pi}\left[-\frac{x_m-x_j}{(x_m-x_j)^2+(y_m-y_j)^2} + \frac{x_m-x_j}{(x_m-x_j)^2+(y_m+y_j)^2}\right],\label{eq:dydt}
\end{align}
respectively. 

Due to the nonlinear terms and the mutual coupling, analytical solutions are not available for this system and the dynamics that arise are very complicated. To analyze the vortex pair motions in the $(x,y)$-plane, we numerically integrate equations~\eqref{eq:dxdt} and \eqref{eq:dydt} using a fourth-order Runge-Kutta scheme, with initial positions for each vortex point given by $(x_j(0),y_j(0))=(x_j^0,y_j^0)$. Therefore, the vortex dynamics is completely specified given the initial position~$x_j^0$ along the $x$-axis, the initial size of the ring (ring diameter is equal to $2y_j^0$), and its strength~$K_j$. 
The time step was selected as $\Delta t = 0.02$ following a brief analysis in which the time step was systematically decreased until the results became independent of the chosen value.

\section{Interaction of two coaxial vortex pairs}\label{sec:twoRings}

Before characterizing the three-vortex-pair case, 
we analyze the motion of two vortex pairs to gain important insights into the interactions of multiple vortex pairs.
This problem was initially studied by Helmholtz~\cite{helmholtz1858integrale}, who pointed out the existence of the leapfrogging motion of vortex pairs. Since then, there have been many studies investigating in detail different aspects of this phenomenon~\cite{behring2019stability,grobli1877specielle,hicks1922mutual,shashikanth2003leapfrogging}. Vortex pairs do not always leapfrog, but an explicit criterion for periodic leapfrogging motion to occur in terms of the initial horizontal separation of the two coaxial vortex pairs ($d=x_2^0-x_1^0$) has been determined in Mavroyiakoumou and Berkshire~\cite{mavroyiakoumou2020collinear}. The leapfrogging criterion is given as a function of the vortex pair strengths and their initial sizes, and it takes the form:
\begin{equation}\label{eq:berkshire}
     d^2<4y_1^0y_2^0\left[1-\left(\frac{K_1^2+K_2^2}{K_1y_1^0+K_2y_2^0}\right)^{(K_1/K_2)+(K_2/K_1)}\left(\frac{y_1^0}{K_1}\right)^{K_1/K_2}\left(\frac{y_2^0}{K_2}\right)^{K_2/K_1}\right]^{-1}-(y_1^0+y_2^0)^2.
\end{equation}
This explicit criterion is in general agreement with the results obtained by a different analysis in Eckhardt and Aref \cite[App.\ B]{eckhardt1988integrable}.
We non-dimensionalize equation~\eqref{eq:berkshire} by dividing both sides by $y_1^0y_2^0$, which results in:
\begin{equation}\label{eq:berkshireDimless}
    D^2<4\left[1-\left(\frac{1/\mu+\mu}{1/\Upsilon+\mu}\right)^{\mu}-\left(\frac{1/\mu+\mu}{1/\mu+\Upsilon}\right)^{1/\mu}\right]^{-1}-\left(\frac{1}{\Upsilon}+2+\Upsilon \right)=:\ell,
\end{equation}
where $D^2=d^2/(y_1^0y_2^0)$, $\Upsilon:=y_2^0/y_1^0$ and $\mu:=K_2/K_1$.
We note that each vortex pair $j=1,2$ alone would translate with velocity $K_j/(4\pi y_j^0)$. Intuitively, equation~\eqref{eq:berkshireDimless}~means that if the vortex pairs start at a distance greater than $D$ apart, then no leapfrogging takes place and their separation increases to infinity with or without an overtake. 

\begin{figure}[htpb!]
    \centering
    \includegraphics[width=.6\textwidth]{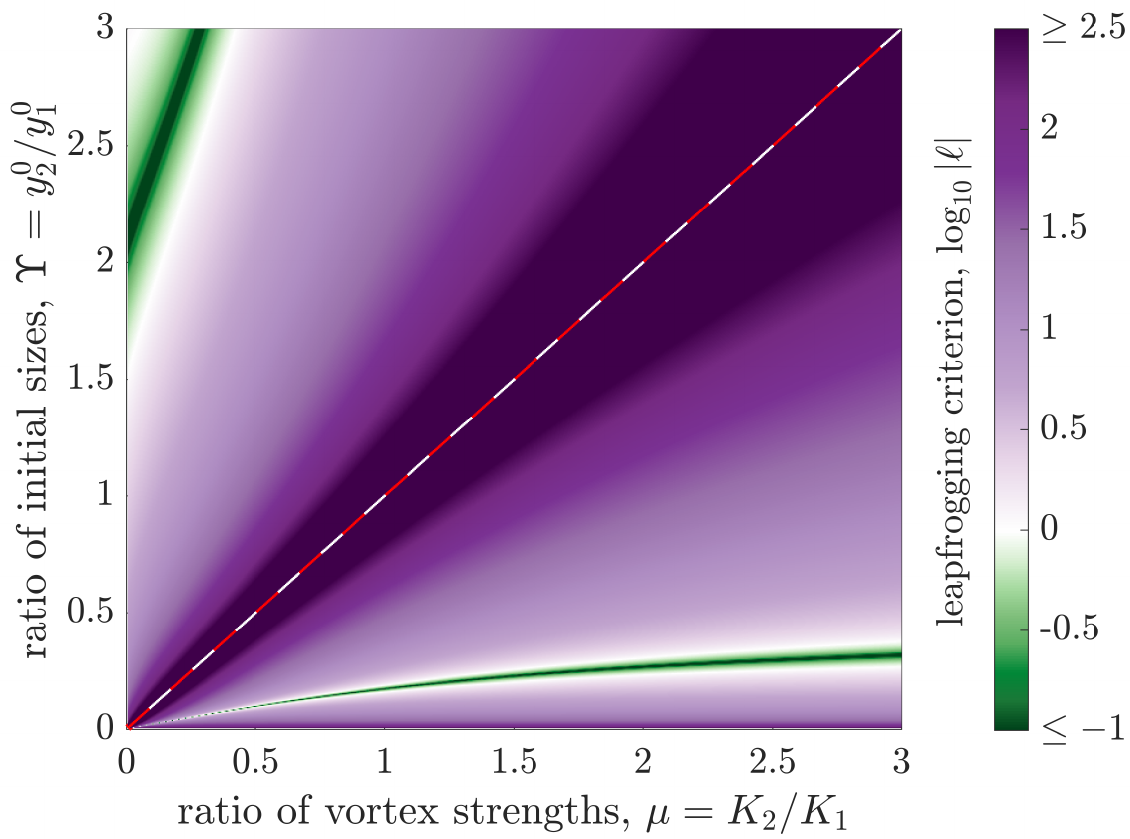}
    \caption{The background colors represent  $\log_{10}|\ell|$, where $\ell$ is as defined in equation~\eqref{eq:berkshireDimless}, in the space of $\mu=K_2/K_1$ and $\Upsilon=y_2^0/y_1^0$. The red dashed line shows that two vortex rings will leapfrog  for any $\ell$ when~$\Upsilon=\mu$.}
    \label{fig:surfBerkshire}
\end{figure}

The leapfrogging criterion is visualized in $\mu$-$\Upsilon$ space in Fig.~\ref{fig:surfBerkshire}, with $\log_{10}|\ell|$ represented as the background color. When $\Upsilon = \mu$, the two vortex pairs always leapfrog, regardless of their initial separation, as shown by the red dashed line. The dark green regions (small $\ell$) indicate that for certain $(\mu, \Upsilon)$ combinations, the vortex pairs must start close together for leapfrogging to occur. In contrast, the dark purple regions (large $\ell$) show that for other $(\mu, \Upsilon)$ combinations, the pairs can start far apart and still leapfrog. This typically occurs when $\Upsilon \approx \mu$, with the width of the dark-purple region increasing as $\mu$ grows. This suggests that as the strength of vortex pair 2 increases relative to pair 1, a greater initial size discrepancy is allowed, and the pairs can still leapfrog despite differences in their self-induced velocities.

\begin{figure}[htpb!]
    \centering
     \includegraphics[width=\textwidth]{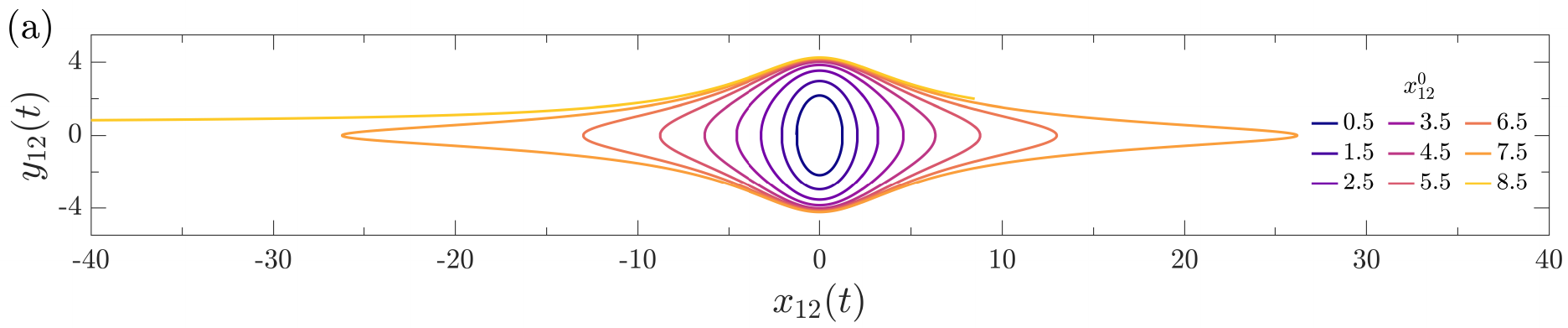}
    \includegraphics[width=\textwidth]{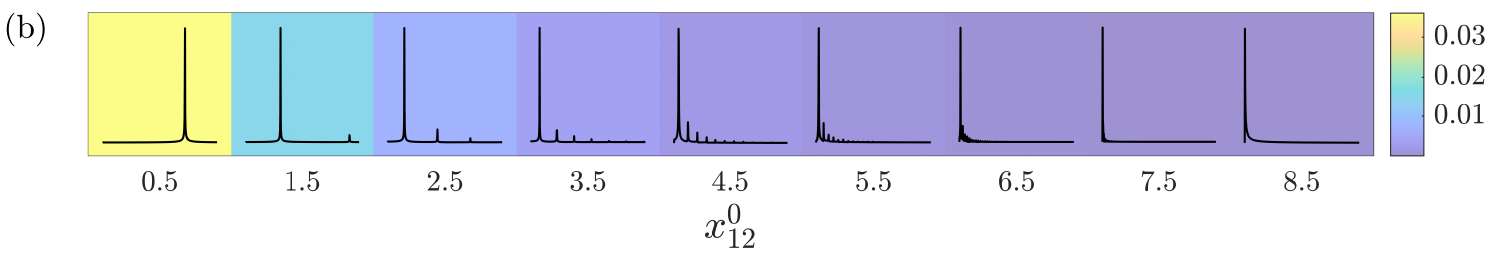}
    \caption{Interaction of two vortex pairs with equal strength ($K_1=K_2=1$) but with different initial sizes ($y_1^0=1$, $y_2^0=2$). (a) $y_{12}(t)$ versus $x_{12}(t)$ for various initial conditions in terms of horizontal spacing between the two vortex rings, ranging from $x_{12}^0=0.5$ to 8.5 in increments of 0.5. Closed orbits correspond to leapfrogging motions whereas the open trajectory implies overtake. (b) The dominant frequencies of the steady state motions (as colored background) for the nine values of $x_{12}^0$ considered in panel (a). The corresponding power spectra for the dynamics arising from each initial condition are plotted in black. Each power spectrum is a plot of power density (per unit frequency) versus frequency. The axis scales are omitted due to space constraints but our focus here is on the qualitative features only.}
    \label{fig:twoRingsResults}
\end{figure}

We now focus on two vortex pairs with the same strength $K_1=K_2=1$ but of different initial sizes $y_1^0=1$ and~$y_2^0=2$, initially located at $x_1^0=0$ and $x_2^0=x_1^0+x_{12}^0$, respectively, with $x_{12}^0\in\{0.5,1.5,\dots,8.5\}$. In Fig.~\ref{fig:surfBerkshire} this would correspond to the point $(\mu,\Upsilon)=(1,2)$. With these values of $(y_1^0,y_2^0,K_1,K_2)$, the leapfrogging criterion (equation~\eqref{eq:berkshire}) gives $d\lesssim 7.94$. Therefore, in Fig.~\ref{fig:twoRingsResults}(a) all cases with $x_{12}^0\leq 7.5$ form closed orbits in $x_{12}(t)$-$y_{12}(t)$
space. When the initial separation between the two vortex pairs is larger than the critical $d$, e.g.\ $x_{12}^0=8.5$, then the first ring moves to $+\infty$ under the influence of its self-induced velocity, and $x_{12}(t)\to -\infty$; depicted as an open trajectory in $x_{12}(t)$-$y_{12}(t)$ space. 
The temporal dynamics corresponding to the vortex pair motions are quantified by computing the power spectra of time series of the relative difference between the $x$-coordinates of the two vortex rings, $x_{12}(t)$. In Fig.~\ref{fig:twoRingsResults}(b) we show these spectra, computed using the discrete Fourier transform (DFT)~\cite{marple1987digital}, across the same range of $x_{12}^0$ values considered in panel (a). The colors denote the dominant frequencies of the steady state motions. As $x_{12}^0$ increases, the dominant frequency decreases.
Essentially each power spectrum has one peak, corresponding to a single frequency, which implies that in the case of the interaction between two vortex pairs the motion is always regular, independent of the initial conditions. 

Previous work~\cite{love1893motion,behring2019stability,grobli1877specielle,tophoj2013instability} has computed the leapfrogging period $T_{\mathrm{leapfrog}}$ for two vortex pairs of the same strength ($K_1=K_2=1$). In all these studies the initial configuration is the same:  $x_{12}^0=0$ and $y_2^0/y_1^0=\alpha/ 1=\alpha$, where $\alpha\in((3-\sqrt{5})/2,1)$---the permissible range for leapfrogging motion~\cite{acheson2000instability}. Their formulas for $T_{\mathrm{leapfrog}}$ have different forms because different parameters are used to express the initial conditions. These have neither been confirmed numerically nor compared systematically between them.
Here we follow the same notation as in Behring and Goodman~\cite{behring2019stability} and the procedure  described in Behring~\cite[App.~B]{behring2020dances}. The leapfrogging period $T_{\mathrm{leapfrog}}$ is a function of the initial positions of the vortex pairs, here given explicitly for
\textit{any} initial horizontal separation $x_{12}^0$ (not just $x_{12}^0=0$) but within the permissible range for leapfrogging, i.e.\ $x_{12}^0$ satisfying equation~\eqref{eq:berkshire}.

To derive an explicit formula for the leapfrogging period, it is convenient to formulate the interacting point vortex equations given in~\eqref{eq:dxdt}--\eqref{eq:dydt} as a Hamiltonian system~\cite{aref2007point,aref2010150,tophoj2013instability}:
\begin{equation}\label{eq:logHamiltonian}
    H(X,Y)=-\frac{1}{2\pi}\log\left(\frac{1}{1-Y^2}-\frac{1}{1+X^2} \right),
\end{equation}
where $X=(x_1(t)-x_2(t))/\hat{d}$ and $Y=(y_1(t)-y_2(t))/\hat{d}$, with $\hat{d}=y_1(0)+y_2(0)$. 
Using logarithmic rules we can rewrite equation~\eqref{eq:logHamiltonian} as 
\begin{equation}
    \frac{(1+X^2)(1-Y^2)}{X^2+Y^2}=h=e^{2\pi H}.
\end{equation}
For this curve to be closed, which corresponds to leapfrogging motion, it must be possible to solve for $X$ when $Y=0$. This gives level curves $1<h<\infty$. 
Using the same initial conditions as in Toph{\o}j and Aref~\cite{tophoj2013instability} and Acheson~\cite{acheson2000instability}, namely starting with the vortex pairs at $(x_1(0),\pm y_1(0))=(0,\pm 1)$ and at $(x_2(0),\pm y_2(0))=(0,\pm \alpha)$ with $\alpha<1$, we have:
\begin{equation}\label{eq:ICsHamiltonian}
    X(0)=X_0=0\quad ; \quad Y(0)=Y_0=\frac{y_1(0)-y_2(0)}{\hat{d}}=\frac{1-\alpha}{1+\alpha}=\frac{1}{\sqrt{1+h}}.
\end{equation}
Solving equation~\eqref{eq:ICsHamiltonian} for $\alpha$ when $h=1$ (the lower limit) we obtain $\alpha=3-2\sqrt{2}$. Thus the permissible range for leapfrogging is $\alpha\in(3-2\sqrt{2},1)$, although \textit{stable} leapfrogging occurs when $\alpha\in((3-\sqrt{5})/2,1)$~\cite{acheson2000instability}.

To compute the leapfrogging period for arbitrary initial horizontal separations between the two vortex pairs, $X(0)$ in equation~\eqref{eq:ICsHamiltonian} must be modified to the general form
\begin{equation}
    X(0)=X_0=\frac{x_1(0)-x_2(0)}{\hat{d}}.
\end{equation}
This corresponds to a different level curve of $h$ (see equation~\eqref{fig:HamiltonianLevelCurves} in \ref{app:Tleapfrog}) whose value is given by:
\begin{equation}\label{eq:hInTleapfrog}
    h=e^{2\pi H_0} \quad ; \quad H_0=-\frac{1}{2\pi}\log\left(\frac{1}{1-Y_0^2}-\frac{1}{1+X_0^2}\right);
\end{equation}
having substituted $(X,Y)=(X_0,Y_0)$ in equation~\eqref{eq:logHamiltonian}. The period of the leapfrogging motion is
\begin{equation}\label{eq:Tleapfrog}
      T_{\mathrm{leapfrog}}(h) = \frac{\pi}{h} \left[\frac{8h^4}{h^2-1}E\left(\frac{1}{h} \right) - 8h^2K\left(\frac{1}{h}\right)\right],
\end{equation}
where $h$ is as in equation~\eqref{eq:hInTleapfrog}, and $E$ and $K$ are the complete elliptic integrals of the first and second~kind, respectively.
The details for the derivation of $T_{\mathrm{leapfrog}}$ have been included in~\ref{app:Tleapfrog}.

\begin{figure}[htpb]
    \centering
    \includegraphics[width=.52\textwidth]{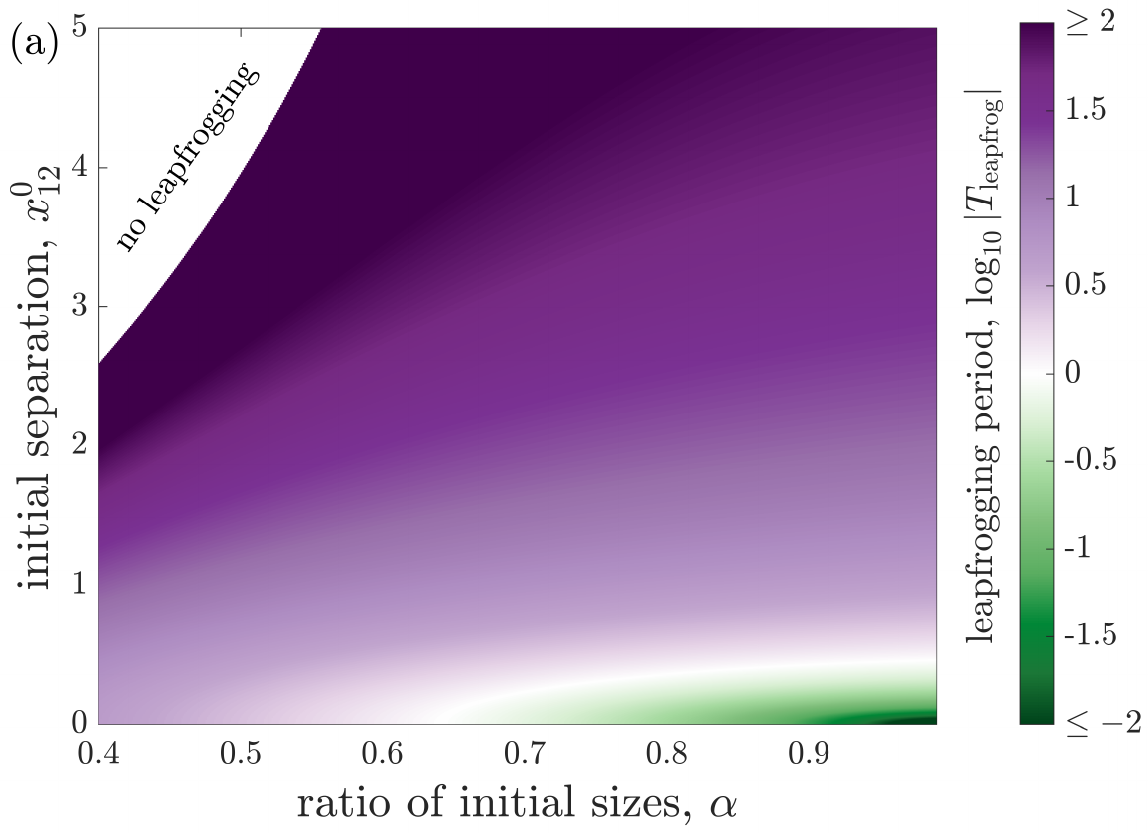}
        \hfill\includegraphics[width=.47\textwidth]{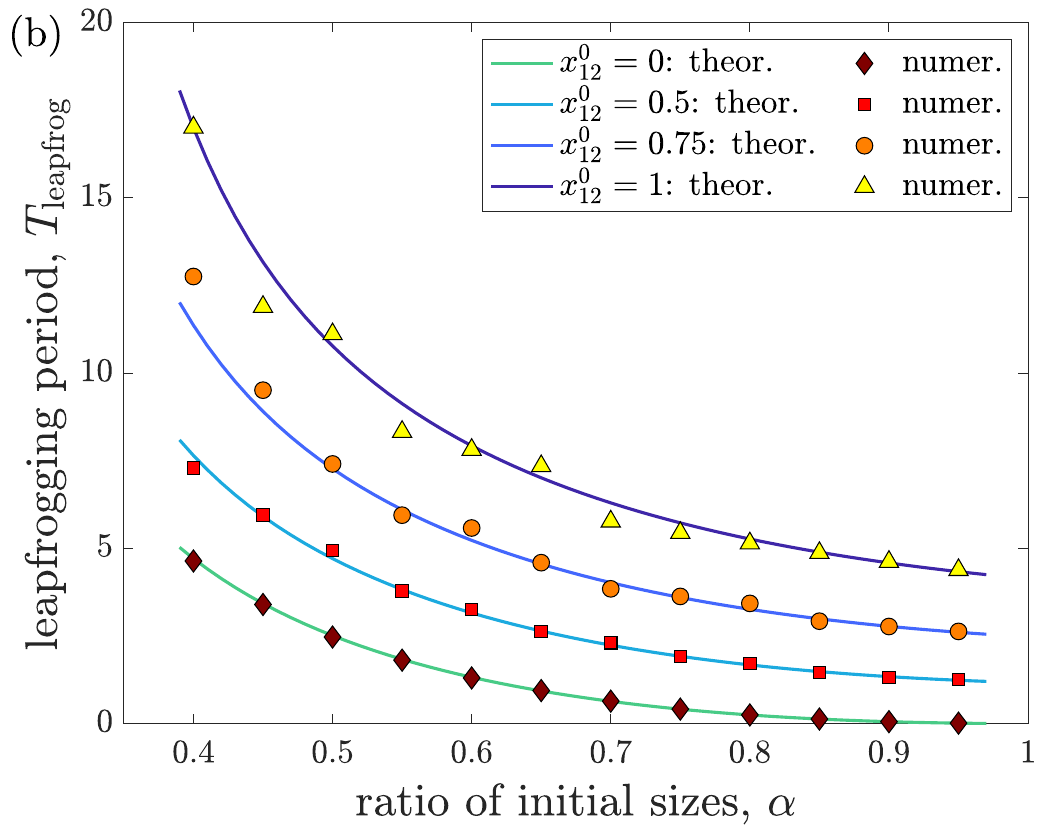}
    \caption{(a) The background colors represent $\log_{10}|T_{\mathrm{leapfrog}}|$, where $T_{\mathrm{leapfrog}}$ is as defined in equation~\eqref{eq:Tleapfrog}. The white region in the top left corner represents the $\alpha$-$x_{12}^0$ regime where the two vortex pairs do not leapfrog, and so there, the leapfrogging period is not defined. (b) Comparison between the numerically-computed (red-shaded markers) and the theoretically-predicted (solid blue-shaded lines) leapfrogging period $T_{\mathrm{leapfrog}}$ versus $\alpha$ for four values of initial horizontal separation~$x_{12}^0$.}
    \label{fig:Tleapfrog}
\end{figure}
We show in Fig.~\ref{fig:Tleapfrog}(a) as colored background the leapfrogging period, $\log_{10}|T_{\mathrm{leapfrog}}|$, in $\alpha$-$x_{12}^0$ space, with $T_{\mathrm{leapfrog}}$ as given analytically in equation~\eqref{eq:Tleapfrog}. The white background at large $x_{12}^0$ and small $\alpha$ indicates the no-leapfrogging region. 
The leapfrogging period is smallest at small $x_{12}^0$ and large $\alpha$ (green region), where the initial size of the vortex pairs is comparable and when they initially in close proximity. The leapfrogging period increases as $x_{12}^0$ increases and $\alpha$ decreases, as the two vortex pairs interact more weakly in that case. For four different initial horizontal separations $x_{12}^0\in\{0,0.5,0.75,1\}$, we compare in Fig.~\ref{fig:Tleapfrog}(b) the theoretically-predicted leapfrogging period (equation~\eqref{eq:Tleapfrog}; solid blue-shaded lines) with the numerically-computed leapfrogging period (red-shaded markers: diamonds, squares, circles, and triangles at $x_{12}^0=0,0.5,0.75$, and 1, respectively). We find excellent agreement between the two, especially at the smaller values of~$x_{12}^0$ where they coincide even at small values of~$\alpha$. The trends for how the leapfrogging period depends on $\alpha$ and~$x_{12}^0$ are more evident in Fig.~\ref{fig:Tleapfrog}(b). The leapfrogging period is largest at the smallest $\alpha$ and follows a decreasing trend with increasing $\alpha$ in all cases. 
When one vortex pair is significantly larger than the other, the larger pair has a slower propagation speed due to its increased size. This slower propagation affects the time it takes for the vortex pairs to interact and exchange positions, resulting in a longer leapfrogging period.

We use these values of $(y_1^0,y_2^0,K_1,K_2)$ as our baseline case in Section~\ref{sec:threeRings} when we consider the interaction between three coaxial vortex pairs. The influence of additional coaxial vortex pairs could be considered with the current methodology but we defer this analysis to future work. An example of the intricate dynamics that can be observed in the case of four interacting vortex pairs is presented in~\ref{app:fourPairs}.

\section{Interactions of three coaxial vortex pairs} \label{sec:threeRings}

We introduce an additional vortex pair to the system, moving along the same axis and in the same direction as the other two. The vortex pairs are labeled 1, 2, and 3 according to their initial configuration. With $N=3$ the number of independent parameters increases from three (for $N=2$) to six, allowing the system to potentially exhibit chaotic behavior~\cite{blackmore2000transition,ting1998asymptotic}, as opposed to the regular motion described in Section~\ref{sec:twoRings}. We focus on a baseline case with $K_1=K_2=1$, $y_1^0=1$, and $y_2^0=2$, while varying $K_3$, $y_3^0$, and the initial horizontal separations $x_{12}^0$ and $x_{23}^0$, between pairs 1 and 2 and between pairs 2 and 3, respectively. 

\subsection{Classification of motions}

\begin{figure}[htpb!]
    \centering
    \includegraphics[width=.95\textwidth]{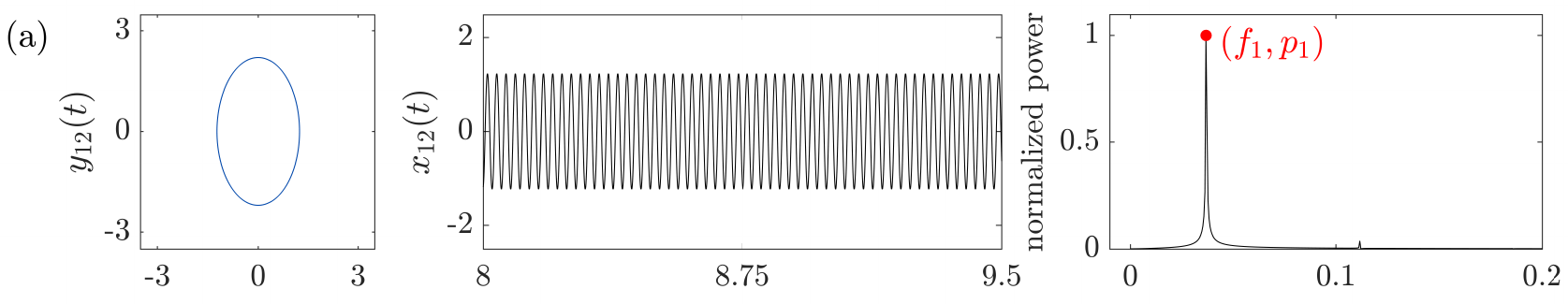}
    \includegraphics[width=.95\textwidth]{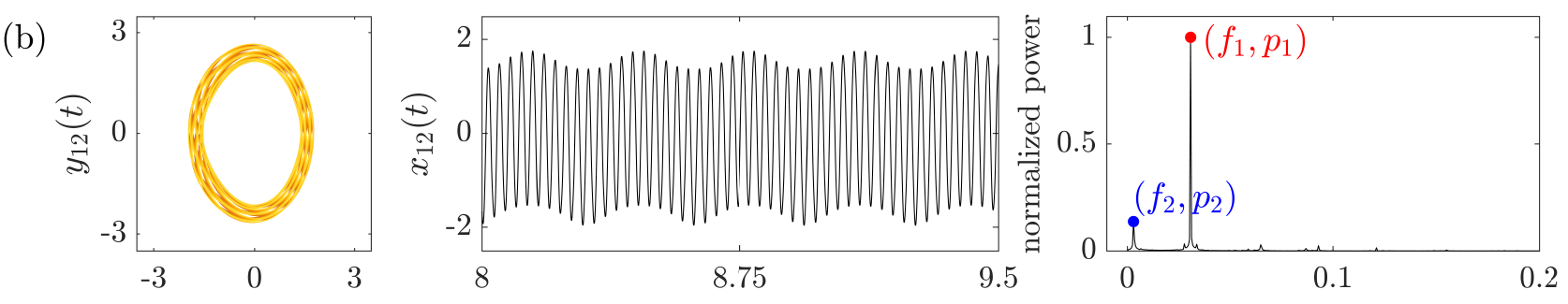}
    \includegraphics[width=.95\textwidth]{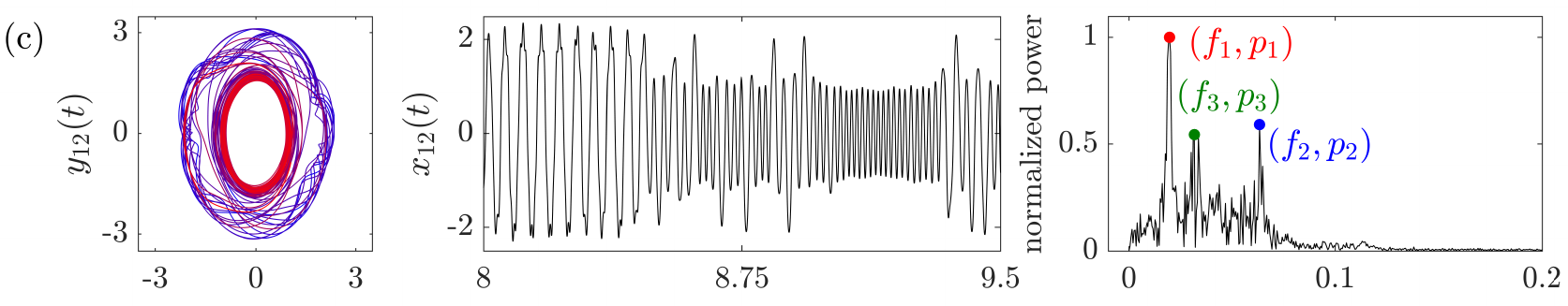}
    \includegraphics[width=.95\textwidth]{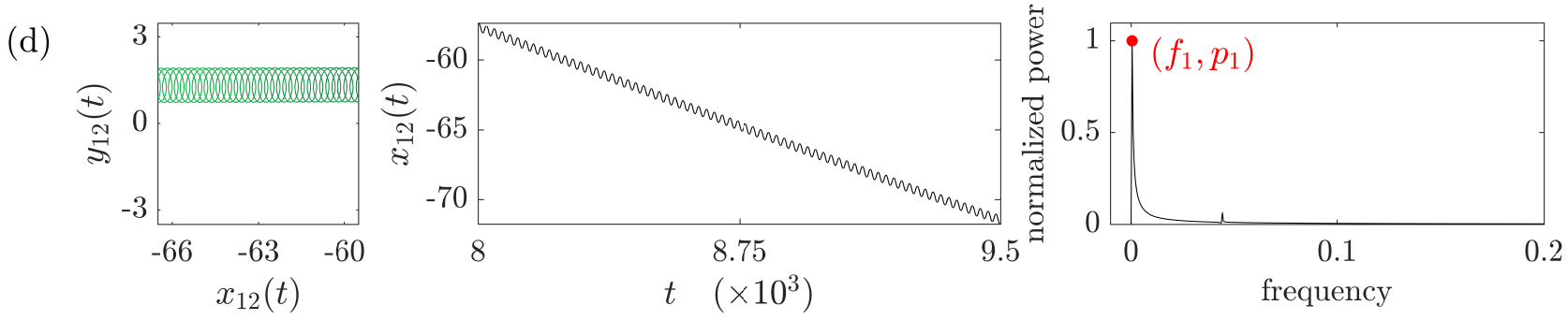}
    \caption{The four types of motions  observed in the interaction of three coaxial vortex pairs. We fix the strength of vortex pair 3 as $K_3=0.5$ and $(y_3^0,x_{12}^0,x_{23}^0)$ equal to (a) $(5/2,0.5,4.5)$ [periodic], (b) $(5/4,0.5,1.5)$ [quasi-periodic], (c) $(1/2,0.5,0.5)$ [chaotic], and (d) $(5/6,4.5,0.5)$ [overtake]. Left column: $y_{12}(t)$ vs $x_{12}(t)$ phase plot. Different colors are used to indicate the four types of motions. (a) Blue is used for periodic leapfrogging, (b) yellow to orange is used for quasi-periodic leapfrogging; the darkening shades of the color scheme correspond to the time evolution. (c) Blue to purple to red is used for chaotic motions. (d) Darkening shades of green are used for overtake/no-pass through cases. Middle column: time variations of the horizontal distance between pairs 1 and 2 ($x_{12}(t)$ vs. $t$). Right column: power spectrum of $x_{12}(t)$ normalized by the maximum value for clarity. The dominant frequency and corresponding power are labeled with $(f_1,p_1)$ [red], the second dominant frequency and power with $(f_2,p_2)$ [blue] and the third by $(f_3,p_3)$ [green].
    }
    \label{fig:classifyMotions}
\end{figure}

To characterize the motions we plot the relative phase trajectories $y_{jm}(t)=y_m(t)-y_j(t)$ versus $x_{jm}(t)=x_m(t)-x_j(t)$ with $j<m\in\{1,2,3\}$ for each vortex pair at its steady state, i.e.\ in the large-time limit, using the last 20\% of the computed time-series. From these relative phase trajectories we observe four main types of motions: i.\ periodic leapfrogging, ii.\ quasi-periodic leapfrogging, iii.\ chaotic motions, and iv.\ overtake/no pass-through.

The temporal dynamics corresponding to these motions are quantified by computing the power spectra of time series of $x_{12}(t)$, $x_{13}(t)$, and $x_{23}(t)$.
In Fig.~\ref{fig:classifyMotions} we fix $K_3=0.5$ and show examples of the $y_{12}(t)$ vs $x_{12}(t)$ phase plots (left column), time series of $x_{12}(t)$ (middle column), and power spectra (right column) that correspond to the four different types of motions (i.--iv.). As in Section~\ref{sec:twoRings}, we use the DFT to obtain a power spectrum of each vortex motion; which reveals the various frequency components associated with each interaction. For leapfrogging motions, periodic and quasi-periodic cases, the power spectrum peaks at discrete values, as seen in the third column of panels (a) and (b). For the chaotic motions it is a continuous band of frequencies (third column of panel (c)). In the third column of panel (d) we show a case of overtake. There, the dominant frequency is essentially zero which accounts for the aperiodic motion. We label the first three dominant frequencies and corresponding powers with $(f_i,p_i)$ for $i=1,2,3$.

Distinguishing between the regular (periodic and quasi-periodic) and chaotic interactions requires defining specific criteria using the  associated frequency power spectra. In particular, we use
\begin{equation}
f_1=\underset{f}{\mathrm{argmax}}(p)\quad ;\quad \Delta f = \frac{1}{2}\left(|f_2-f_1|+|f_3-f_2|\right)    \quad ;\quad p_r = \frac{1}{2}\left(\frac{p_1}{p_2}+\frac{p_2}{p_3}\right),
\end{equation}
where $f_1$ corresponds to the dominant frequency, $\Delta f$ is the average distance between the first and second most dominant frequencies $|f_2-f_1|$ and the second and third $|f_3-f_2|$, and $p_r$ can be thought of as a type of harmonic mean between the first three most dominant powers ($p_1,p_2,p_3$).
The criteria are:
\begin{equation}
    \begin{cases}
    \text{Chaotic:}& \Delta f<10^{-2} \quad \text{and} \quad p_r<2,\\
    \text{Periodic:}& \Delta f>2\times10^{-2} \quad \text{and} \quad p_r>2,\\
    \text{Quasi-periodic:} & \text{otherwise}.
    \end{cases}
\end{equation}

Row (a) of Fig.~\ref{fig:classifyMotions} shows an example of a periodic leapfrogging motion with $(x_{12}^0,x_{23}^0)=(0.5,4.5)$ and $y_3^0=5/2$. The relative phase trajectories $y_{12}(t)$ vs $x_{12}(t)$ trace a closed orbit that repeats in a periodic manner. The corresponding $x_{12}(t)$ time series is periodic and the power spectrum contains a dominant sharp peak without other harmonics. Row (b) is with $(x_{12}^0,x_{23}^0)=(0.5,1.5)$ and $y_3^0=5/4$ and the motion is quasi-periodic, still dominated by a single frequency. The power spectrum now has a sequence of sharp peaks that correspond to the short and long periods seen in the $x_{12}$ time series in the middle column. The relative phase trajectories are characterized by quasi-circular orbits that are slightly displaced relative to each other with the color changing from yellow to orange to show the time evolution from early to late times, respectively.
In row (c) the trend toward aperiodicity continues. With $(x_{12}^0,x_{23}^0)=(0.5,0.5)$ and $y_3^0=1/2$ the trajectories (left column) do not have any particular organization and cross their orbits arbitrarily. The time series (middle column) shows peaks with a somewhat regular spacing in certain time intervals  but otherwise chaotic.
The corresponding power spectrum (right) transitions from a discrete one in  panels (a) and (b) to a broadband spectrum, with peaks that are in very close proximity to one another, displaying a gradual decay in the power at higher frequencies, typical of chaotic dynamics. Row (d) shows an overtake case where 1 passes through 2 and then flies off to infinity, indicated by the negative values in $x_{12}(t)$ that tend to negative infinity. The power spectrum reveals that the dominant frequency is zero. 

\begin{figure}[htpb!]
    \centering
    \includegraphics[width=\linewidth]{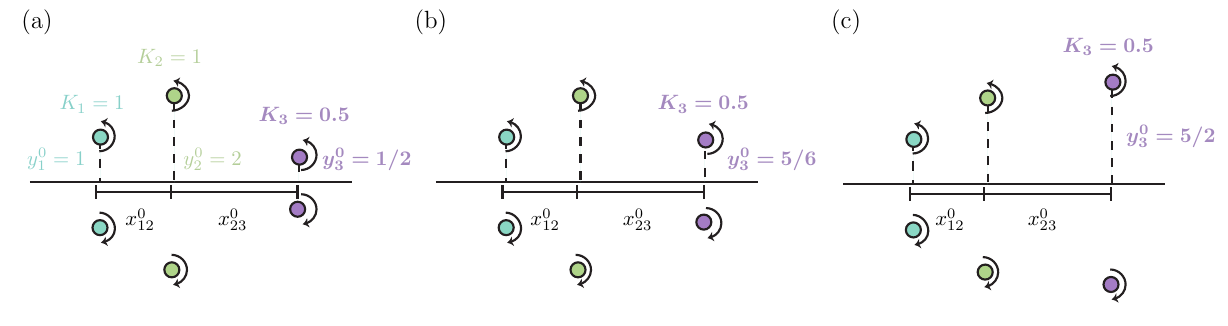}
    \caption{A schematic diagram illustrating the  initial configurations considered in Fig.~\ref{table:plots}. The first two vortex pairs have equal strength $K_1=K_2=1$ while the third vortex pair has half the strength, $K_3=0.5$. The initial sizes of the vortex pairs are $y_1^0=1$, $y_2^0=2$, and $y_3^0$ equal to (a) 1/2, (b) 5/6, (c) 5/2. The initial horizontal separations are in all cases: $x_{12}^0,x_{23}^0\in\{0.5,1.5,2.5,3.5,4.5\}$. }
    \label{fig:schematicThree}
\end{figure}

\begin{figure}[htpb!]
    \centering
    \includegraphics[width=\linewidth]{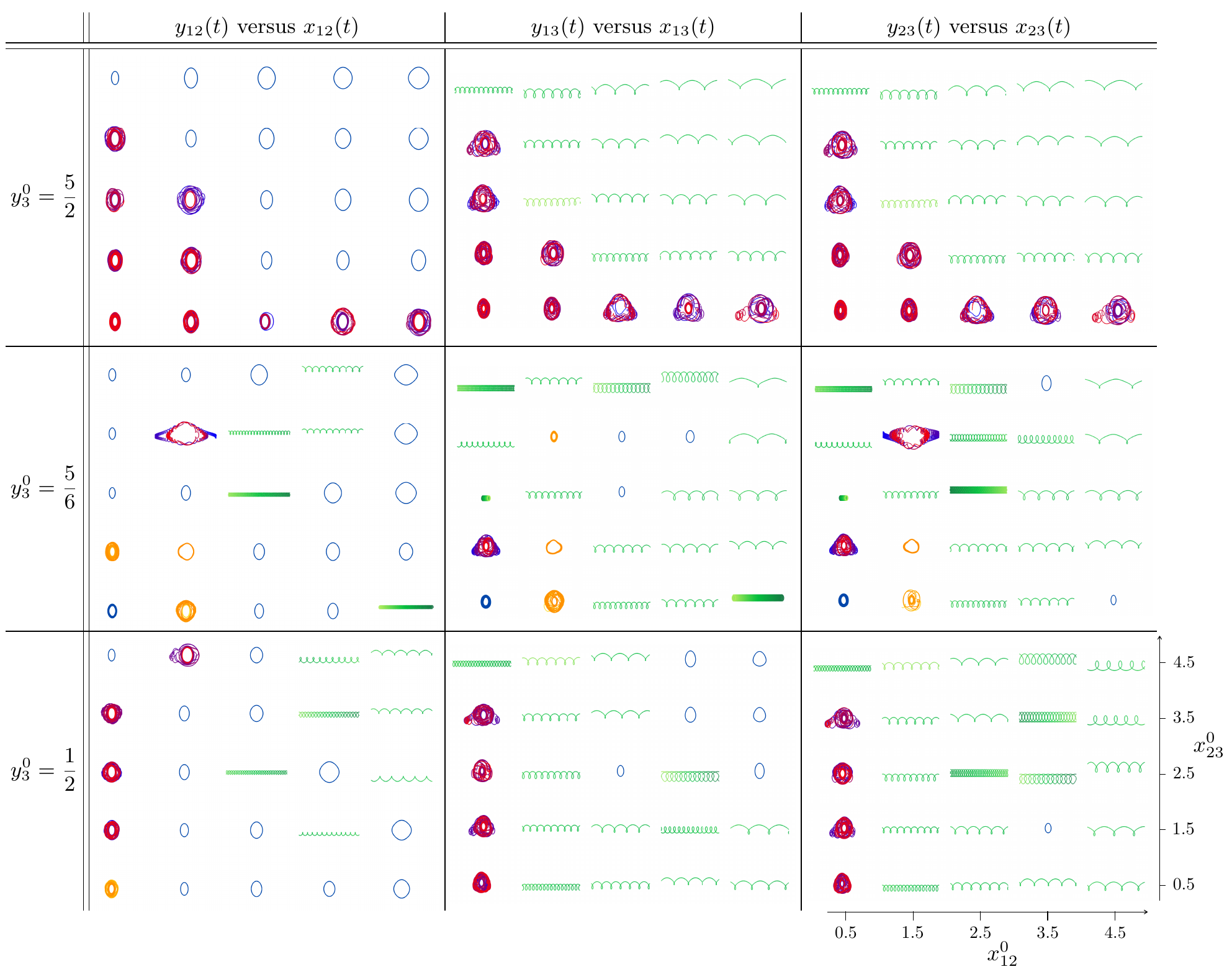}
\caption{Table of plots showing the interactions between vortex pairs 1--2 (column 1), 1--3 (column 2), 2--3 (column~3), for three values of $y_3^0$ (5/2, 5/6, 1/2) and fixed strength $K_3=0.5$. 
Within each subpanel we plot the relative phase trajectories $y_{jm}(t)$ versus $x_{jm}(t)$ with $j<m\in\{1,2,3\}$, for different values of $x_{12}^0$ (from 0.5 to 4.5) and $x_{23}^0$ (also from 0.5 to 4.5), as labeled at the bottom right corner.
We adopt the same color scheme as in Fig.~\ref{fig:classifyMotions} to denote the different motions. Blue is used for periodic leapfrogging. Yellow to orange is used for quasi-periodic leapfrogging; the darkening shades of the color scheme corresponds to the time evolution. Blue to purple to red is used for chaotic motions. Darkening shades of green are used for overtake/no-pass through cases. All cases are in the steady-state, large-time limit. }\label{table:plots}
\end{figure}

We show the wide variety of dynamics that occur in the three-dimensional parameter space $y_3^0$-$x_{12}^0$-$x_{23}^0$ for $K_3=0.5$. The initial configurations and associated pair interactions are shown in Figs.~\ref{fig:schematicThree} and~\ref{table:plots}, respectively.
Fig.~\ref{table:plots} is organized into nine subpanels, each corresponding to a particular value of $y_3^0$ (labeled on the left side), with varying values of $x_{12}^0$ and $x_{23}^0$ within each subpanel, as labeled in the bottom-right corner. The first column of subpanels focuses on the interaction between pairs 1 and 2, the second column on pairs 1 and 3, and the third column on pairs 2 and 3.
For the chosen set of strengths ($K_1=K_2=2K_3=1$), we observe that the interactions are predominantly between the stronger pairs: 1 and 2. 

A comparison of the three columns in Fig.~\ref{table:plots} reveals that periodic leapfrogging (blue color scheme) occurs only between two vortex pairs at a time. For example, for $y_3^0=5/6$, $x_{12}^0=3.5$, and $x_{23}^0=4.5$, vortex pairs 1 and 3 exhibit periodic leapfrogging, while both vortex pairs 1 and 3 overtake pair 2. In such cases, the initial system of three pairs 
effectively disintegrates into two subsystems: a leapfrogging set of vortex pairs and a single vortex pair after some interactions ($3\to 2+1$). 
This behavior is in contrast to quasi-periodic and chaotic dynamics, where the existence of a quasi-periodic or chaotic relative phase trajectory implies similar behavior for the other two relative phase trajectories.
For the quasi-periodic case, such an example is seen with $y_3^0=5/6$, $x_{12}^0=1.5$, and $x_{23}^0=1.5$, and for the chaotic case with $y_3^0=5/2$, $x_{12}^0=0.5$, and $x_{23}^0=3.5$. 
Within each subpanel the following trends emerge: at the smallest $x_{12}^0$ values considered, and for $x_{23}^0\leq 3.5$, the system displays chaotic or quasi-periodic motions, rather than periodic leapfrogging with a single frequency. This occurs because the stronger coupling between all three vortex pairs, prevents periodic behavior. When the three vortex pairs are too close, more complex, chaotic dynamics are observed instead of periodic motion.

At the largest $y_3^0=5/2$ (top row), the third vortex pair is the slowest due to its larger initial size ($y_3^0>y_2^0>y_1^0$; see Fig.~\ref{fig:schematicThree}(c)). The left subpanel shows that pairs 1 and 2 always interact either chaotically or periodically, while in the middle and right subpanels, interactions between 1--3 and 2--3 occur only at the smallest initial separations ($x_{12}^0$ or $x_{23}^0$), and the motions there are chaotic.
As the initial width of the third vortex pair decreases to $y_3^0=5/6$ (middle row), several changes are observed: now the vortex pairs can exhibit quasi-periodic leapfrogging, vortex pair~1 can overtake pair 2 with the two of them eventually stop interacting, and chaotic interactions between the three pairs become less common. Quasi-periodic leapfrogging typically occurs when both $x_{12}^0$ and $x_{23}^0$ are small and cannot co-exist with overtake. As noted earlier, when the initial configuration splits into two interacting vortex pairs and a single vortex pair propagating far away from the others, the two interacting vortex pairs will always exhibit periodic leapfrogging. 

In the middle row of Fig.~\ref{table:plots},  periodic leapfrogging is observed between vortex pairs 1 and~3 for moderate values of $x_{12}^0$ and $x_{23}^0$. This results from weakened interactions between pairs 1 and 2, and between pairs 2 and 3. Since vortex pair 1 has a smaller initial width compared to pair 2, it moves faster and overtakes pair 2. As a result, vortex pair 1 manages to catch up with pair 3, leading to periodic leapfrogging between them. This behavior is also evident at even smaller $y_3^0$ values (bottom row), but it occurs at larger $x_{12}^0$ and $x_{23}^0$. The faster motion of vortex pair 3 weakens its interaction with pair 2, allowing periodic leapfrogging between pairs 1 and~3 only when the interaction between pairs 1 and~2 is sufficiently weak, which occurs when~$x_{12}^0$ is large.

\subsection{Percentage of overtake and periodic leapfrogging motions in parameter space}

\begin{figure}[htpb]
    \centering
    \includegraphics[width=\textwidth]{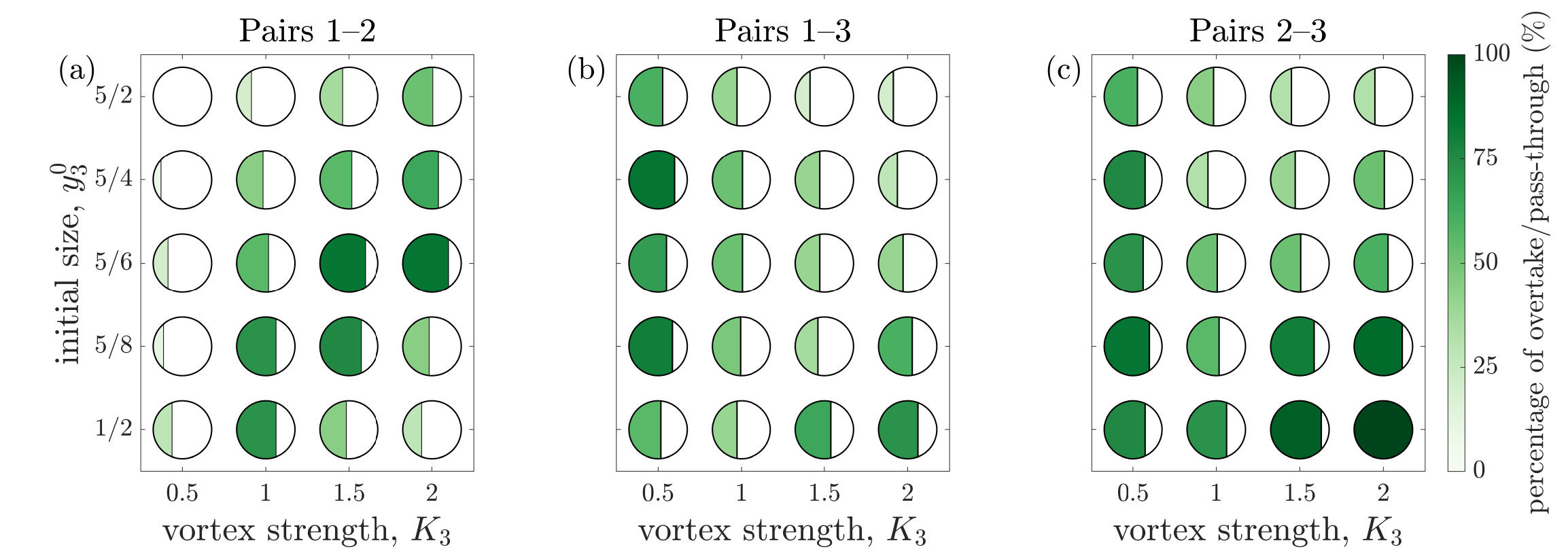}
    \caption{The percentage of overtake/pass-through cases in $K_3$-$y_3^0$ space for the interaction between vortex pairs (a) 1-2, (b) 1-3, and (c) 2--3, taking into account all 25 combinations of initial horizontal separations $x_{12}^0$ and $x_{23}^0$. The colors denote the percentage of overtake or pass-through, ranging from 0\% (white) to 100\% (darkest green). The circles get filled in proportionally to the percentage of overtake/pass-through and with the appropriate color. }
    \label{fig:percentLeapfrog}
\end{figure}

To systematically characterize the dynamics, we expand our parameter space to include four values of strength $K_3$ $(\{0.5,1,1.5,2\})$, five values of initial size $y_3^0$ $(\{1/2,5/8,5/6,5/4,5/2\})$, five values of each of the initial separations $x_{12}^0$ and $x_{23}^0$ $(\{0.5,1.5,2.5,3.5,4.5\})$. This results in 500 distinct cases. For each combination of $K_3$ and $y_3^0$, we calculate the percentage of overtake/pass-through cases across all 25 combinations of initial separations $x_{12}^0$ and $x_{23}^0$, as shown in Fig.~\ref{fig:percentLeapfrog}. The colors represent the percentage of overtake or pass-through, with white indicating 0\% and the darkest green 100\%. The circles get filled in proportionally to the percentage of overtake/pass-through cases and with the appropriate color. 
Panel (a) shows this percentage for the interaction between vortex pairs 1 and 2, panel (b) for 1 and 3, and panel (c) for 2 and 3. 

The first column of each panel corresponds to $K_3=0.5$, which is the vortex strength value analyzed in terms of relative phase trajectories in Fig.~\ref{table:plots}. 
In column 1 of Fig.~\ref{fig:percentLeapfrog}(a) the percentage of overtakes or pass-throughs for vortex pairs 1 and~2 remains below 30\% for all values of $y_3^0$. Notably, in the top left corner, corresponding to the largest $y_3^0$ (5/2), there is $0\%$ overtake between pairs 1 and 2. This occurs because vortex pair 3 is weak and slow, allowing pairs 1 and 2 to overtake it and then leave it behind. In contrast, columns 1 of panels (b) and (c), which show the interactions between 1--3 and 2--3, exhibit a larger percentage of overtake/pass-through cases (50--85\%), with similar percentages across all values of $y_3^0$. This behavior aligns with the low percentages of overtake between pairs 1 and 2.

The case of $K_3=1$ is more complex. With all three vortex pairs having the same strength, the dynamics becomes a competition between the initial sizes of the pairs and the initial separations between them. The interaction between any two vortex pairs is heavily influenced by how they interact with the third pair. As $K_3$ increases to larger values ($>1$), two distinct regimes emerge: one for $y_3^0<5/6$ and another for $y_3^0\geq 5/6$. For large $y_3^0$, increasing $K_3$ results in fewer overtake cases between vortex pairs 1 and 3, and 2 and 3 (upper sections of Figs.~\ref{fig:percentLeapfrog}(b) and~(c)), but more overtake cases between pairs 1 and 2 (upper section of Fig.~\ref{fig:percentLeapfrog}(a)). This occurs because vortex pair 3 travels slowly, allowing vortex pairs 1 and 2 to approach and interact with it. This interrupts the interactions between vortex pairs 1 and 2, which are more likely to now overtake one another.
For smaller $y_3^0$, the trends generally reverse.

\begin{figure}[htpb]
    \centering
    \includegraphics[width=\textwidth]{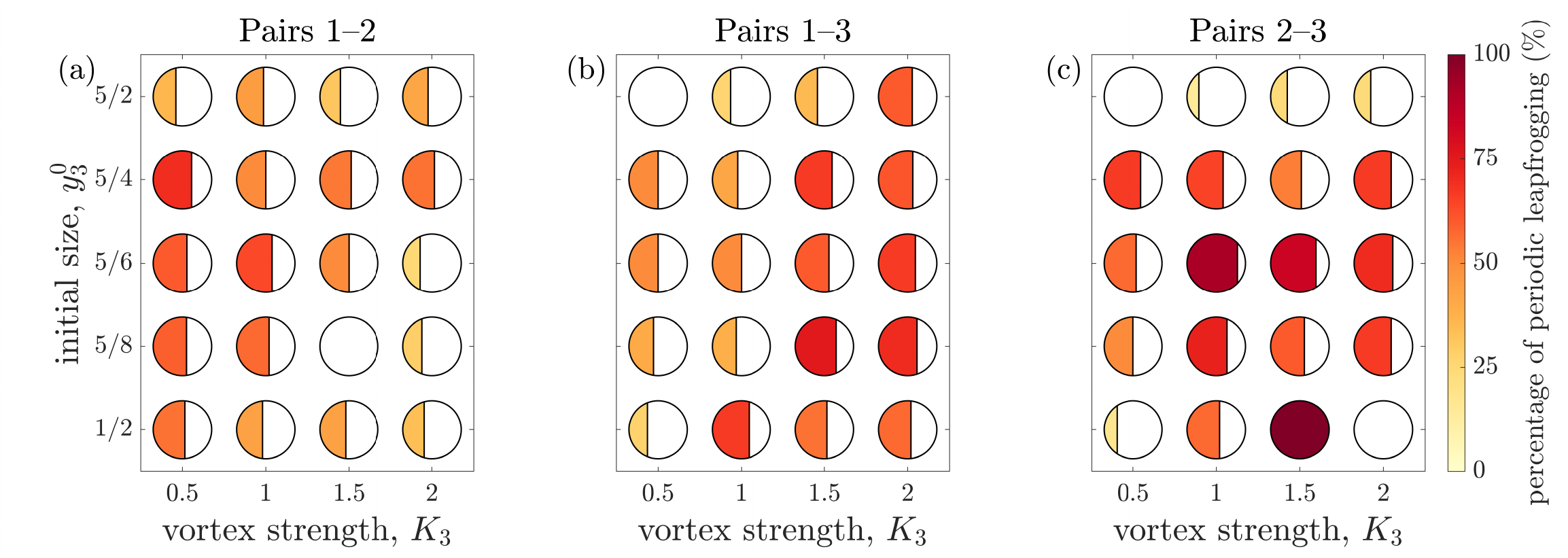}
    \caption{The percentage of \textit{periodic} leapfrogging in $K_3$-$y_3^0$ space for the interaction between vortex pairs (a) 1-2, (b) 1-3, and (c) 2--3, taking into account all 25 combinations of initial horizontal separations $x_{12}^0$ and $x_{23}^0$. The computed values only take into account the total number of interacting cases (including also quasi-periodic leapfrogging and chaotic interactions), therefore excluding the  overtake/pass-through cases.}
    \label{fig:percentPeriodic}
\end{figure}

Excluding the overtake/pass-through cases and focusing on the rest of the interactions: periodic, quasi-periodic, and chaotic, we show in Fig.~\ref{fig:percentPeriodic} the percentage of \textit{periodic} leapfrogging in $K_3$-$y_3^0$ space.
In Fig.~\ref{fig:percentPeriodic}(a), for small $y_3^0$ ($=1/2$) and large $K_3$ ($\geq 1$), 100\% of the interactions between vortex pairs 1 and 2 are periodic leapfrogging. This is due to vortex pair 3 being fast and strong, causing it to move away from the other two pairs. High percentages of periodic leapfrogging are also observed when $K_3=0.5$, as vortex pair 3 is weak in comparison to pairs 1 and 2, this means that the influence on the other two pairs is small, and periodic leapfrogging between pairs 1 and 2 is more likely. As $y_3^0$ increases, the percentage of leapfrogging cases generally decreases, although certain exceptions also exist. For example, at $(K_3,y_3^0)=(1.5,5/6)$ where the percentage of periodic leapfrogging between pairs 1 and 2 is 100\%.

Figure~\ref{fig:percentPeriodic}(b) shows the interaction between vortex pairs 1 and 3. 
Moving from the top left corner (small $K_3$ and large $y_3^0$) to the bottom right corner (large $K_3$ and small $y_3^0$), the percentage of periodic leapfrogging increases monotonically. In the large-$K_3$ and small-$y_3^0$ regime, the three vortex pairs tend to break into two subsystems: overtake (Fig.~\ref{fig:percentLeapfrog}(b)) or high percentages of periodic leapfrogging as shown in Fig.~\ref{fig:percentPeriodic}(b). The trends for the interactions between vortex pairs 2 and 3 are similar to those between pairs 1 and 3. In the upper right corner of Fig.~\ref{fig:percentPeriodic}(c) (when $y_3^0\geq 5/6$), the percentage of periodic leapfrogging increases monotonically with decreasing $y_3^0$, consistent with the increasing number of overtake cases in Fig.~\ref{fig:percentLeapfrog}(c). The trends in the bottom right corner are less obvious in Fig.~\ref{fig:percentPeriodic}(c).

In general, we find that chaotic interactions and quasi-periodic leapfrogging occur when the three vortex pairs have similar strengths and initial size. Otherwise, the system tends to disintegrate into two subsystems: a single propagating vortex pair and two periodically leapfrogging pairs.

\subsection{Dependence of leapfrogging period on parameter space}\label{sec:sensitivity}

\begin{figure}[H]
\centering
\includegraphics[width=.34\textwidth]{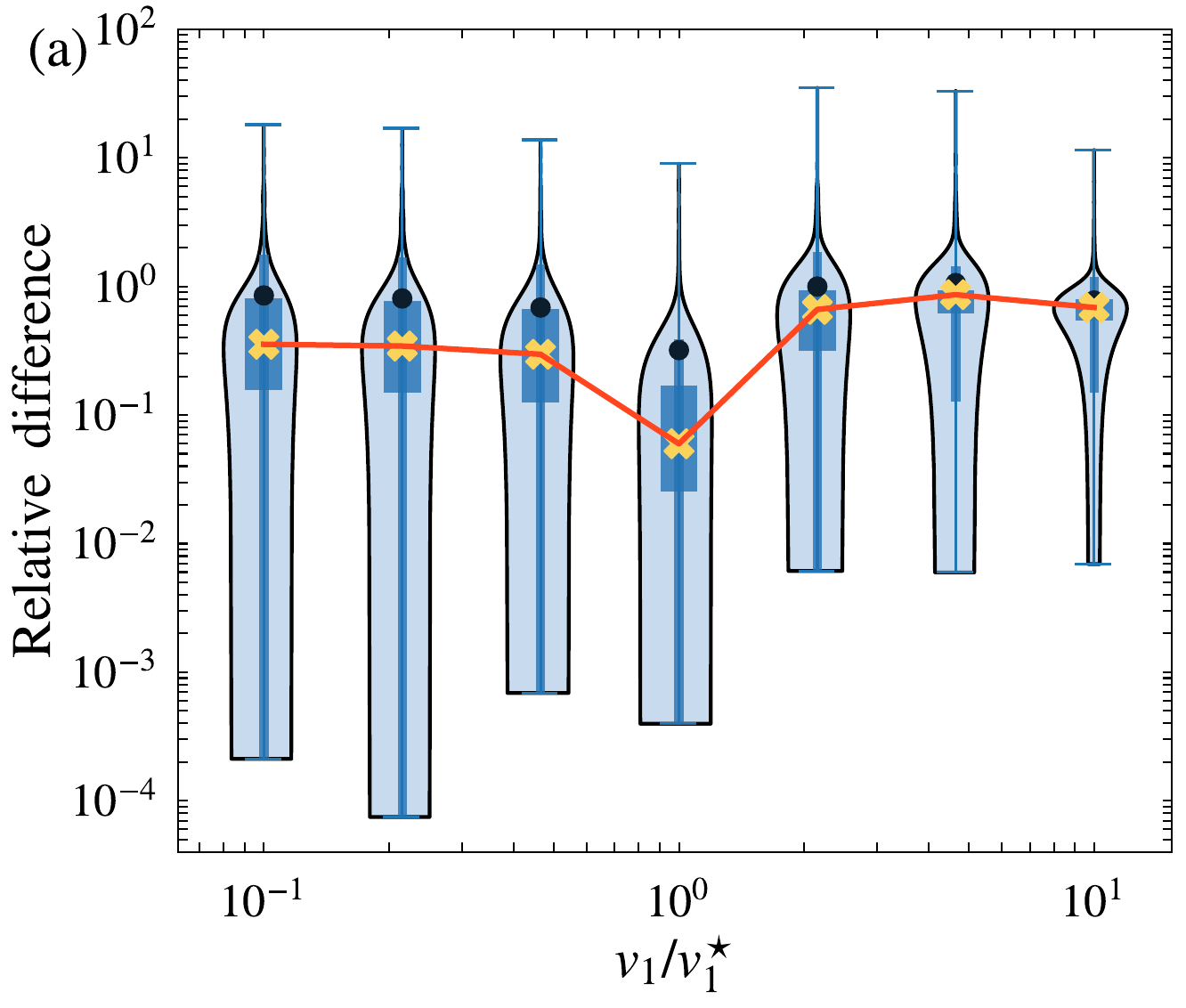}\hfill
\includegraphics[width=.317\textwidth]{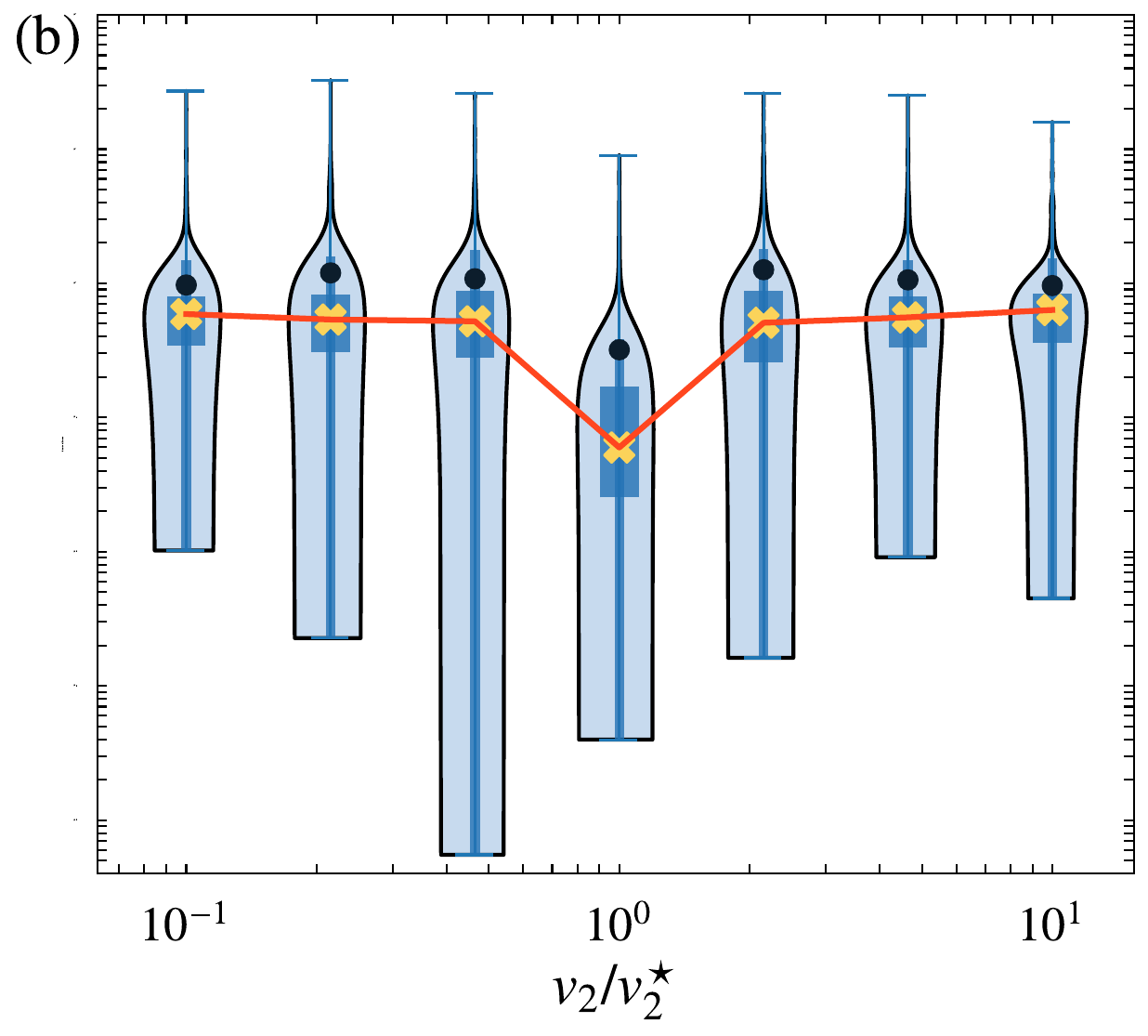}\hfill
\includegraphics[width=.317\textwidth]{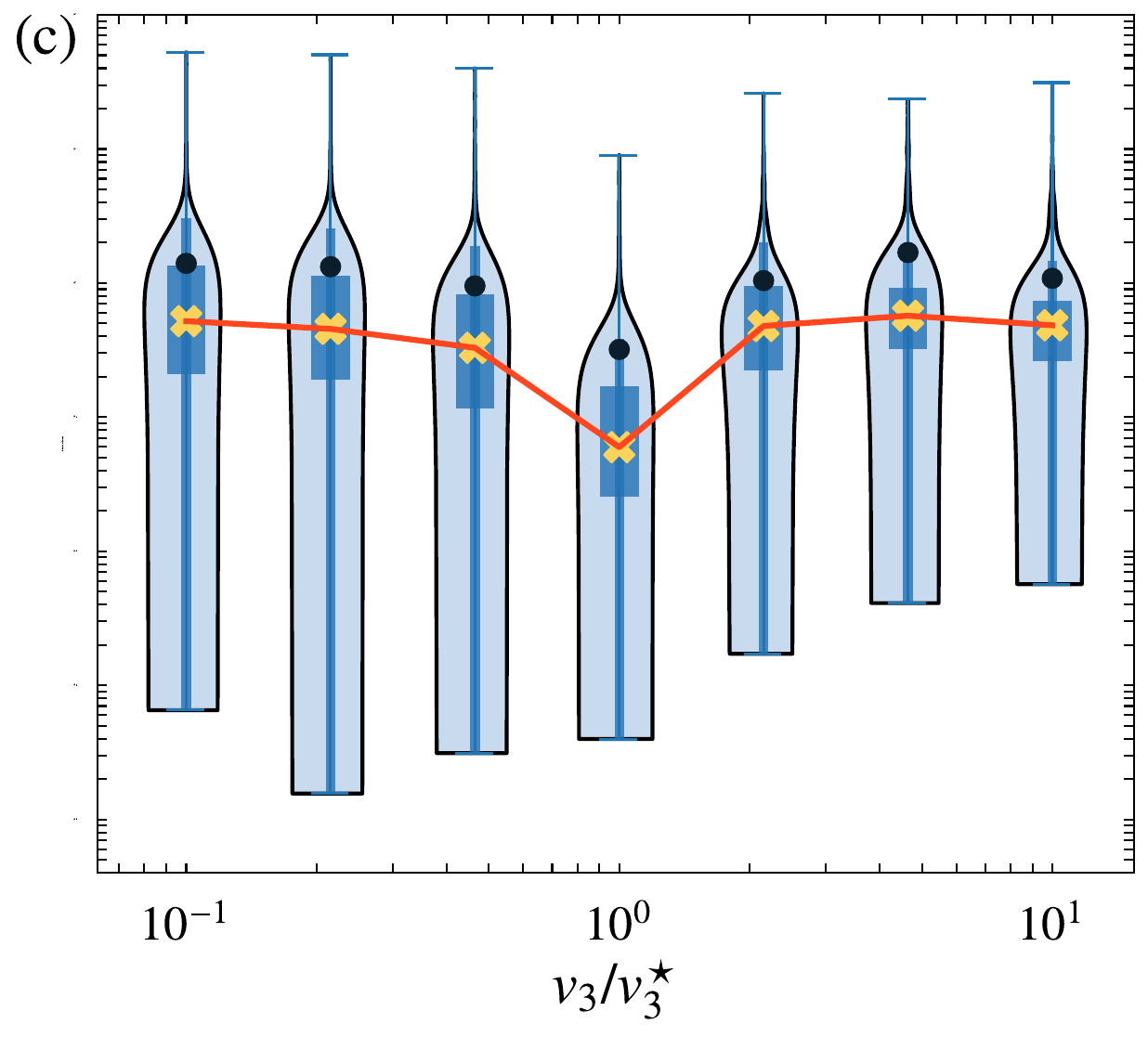}
\includegraphics[width=.34\textwidth]{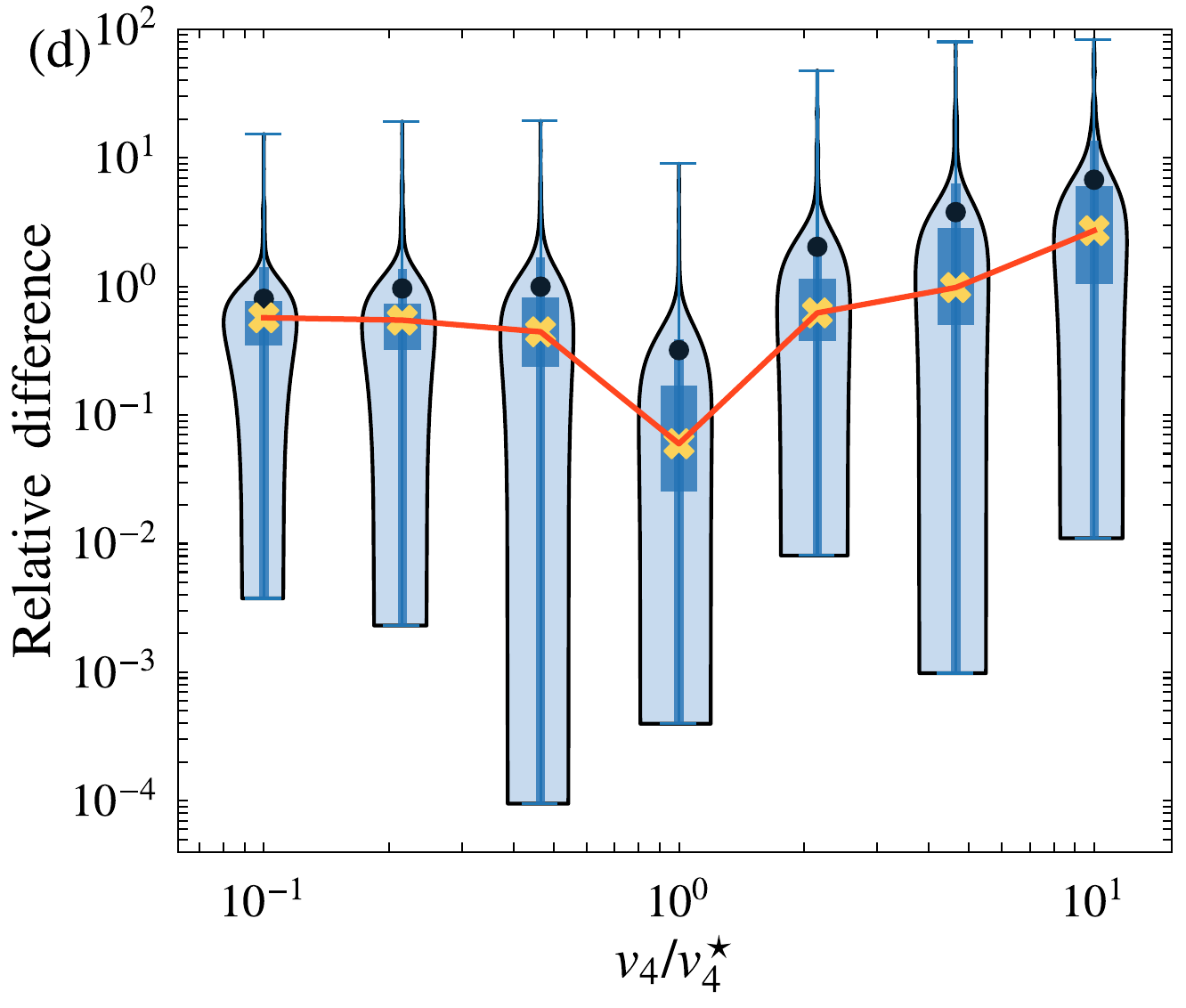}\hfill
\includegraphics[width=.317\textwidth]{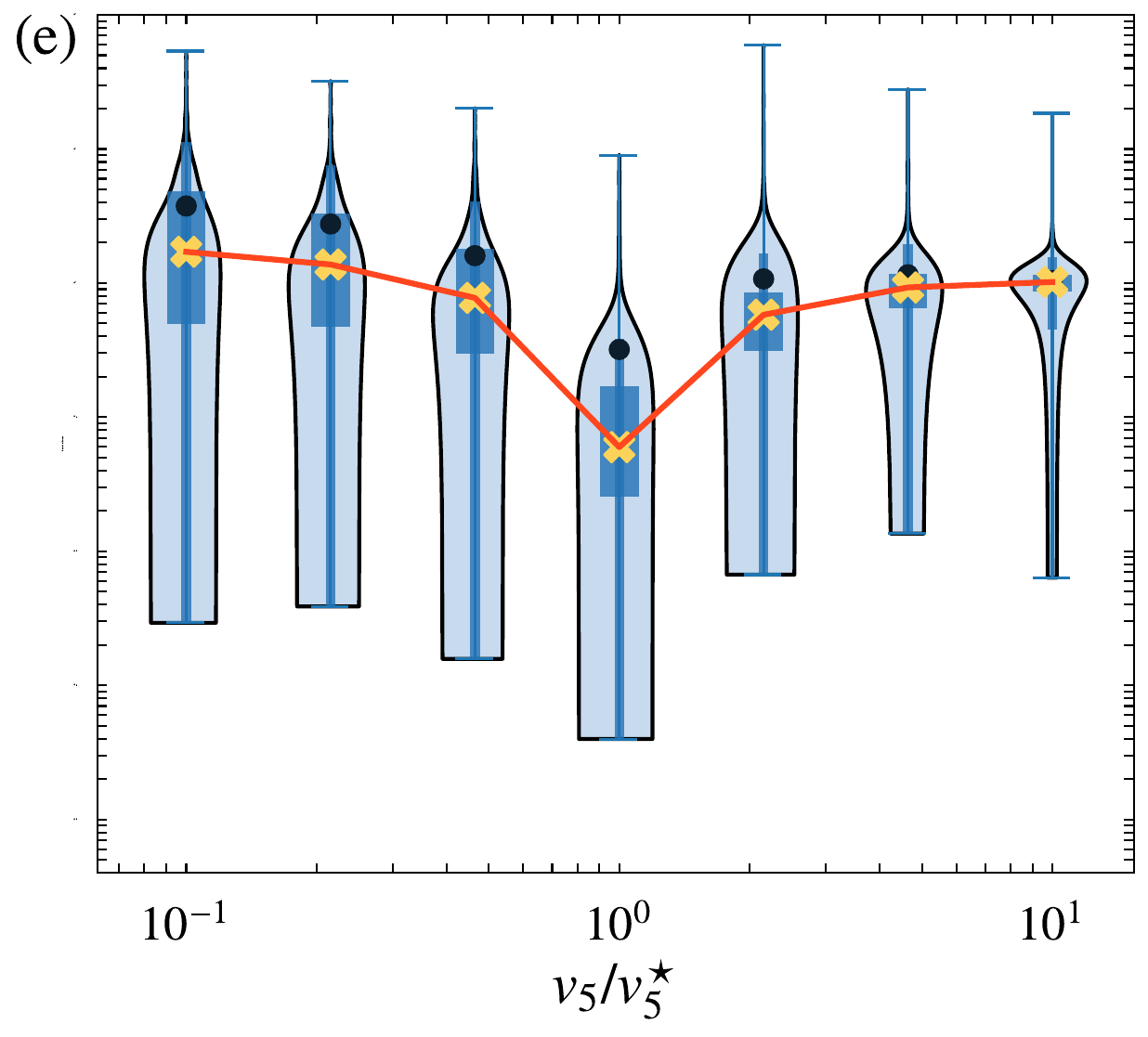}\hfill
\includegraphics[width=.317\textwidth]{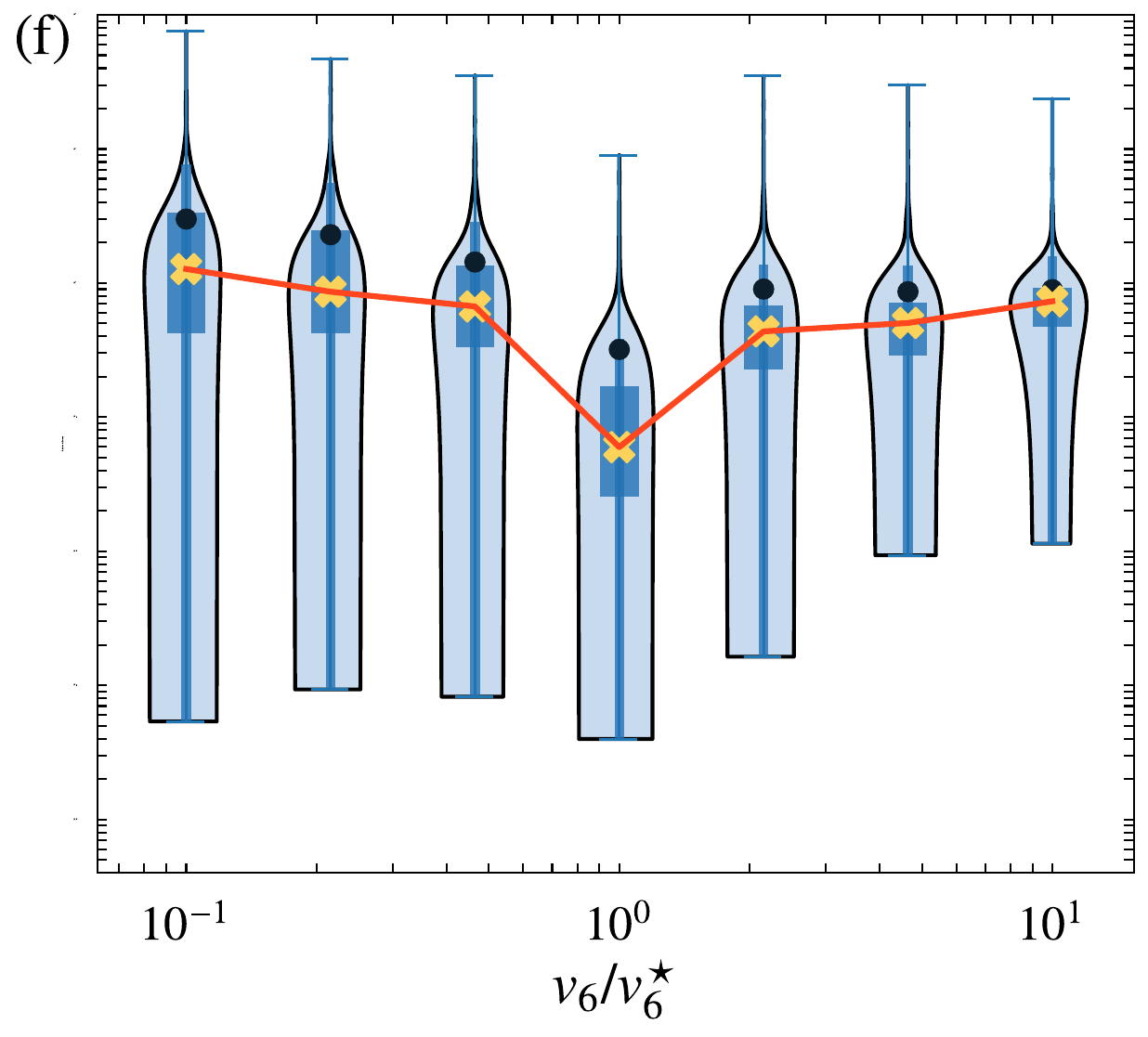}
\caption{Violin plots showing the sensitivity of the leapfrogging period on different physical parameters: (a) $\nu_{1}=y_2^0/y_0$, (b) $\nu_{2}=y_3^0/y_0$, (c) $\nu_{3}=K_2/K_1$, (d) $\nu_{4}=K_3/K_1$, (e) $\nu_{5}=x_{12}^0/y_0$, and (f) $\nu_{6}=x_{23}^0/y_0$, where $y_{0}=(y_{1}^0+y_{2}^0+y_{3}^0)/3$ and $\nu^{\star}$ are the corresponding original values. The relative difference is defined as the difference between the leapfrogging period from the simulation data ($T_{\mathrm{leapfrog}}$) and the model output after varying one parameter, divided by $T_{\mathrm{leapfrog}}$. The violin plots highlight the distribution of the relative differences given a parameter variation. The black dots and the yellow crosses represent the mean and median relative difference values, respectively, and the orange line connects the median values across all the variations. }
\label{fig:sensitivity}
\end{figure}

Now that we have established that periodic leapfrogging is still possible when three coaxial vortex pairs interact, our next objective is to analyze the period of the leapfrogging motion. For the case of two vortex pairs, discussed in Section~\ref{sec:twoRings}, we derived the leapfrogging period analytically using a Hamiltonian framework. However, when extending the system to three coaxial vortex pairs, an analogous analytical derivation of the leapfrogging period proves to be infeasible using the same theoretical approach employed for the two-vortex-pair case.
As previously discussed, the analysis of the interactions between three vortex pairs is a complicated task, not only due to the diversity of their motions but also because the transitions between these different motions are not smooth. Since ``overtake'' events are typically straightforward and chaotic interactions lack a well-defined period, we focus on the periodic leapfrogging motions.
To identify the primary factors influencing the period of leapfrogging motions, we conduct a sensitivity analysis on the dimensionless parameters governing the system. Specifically, we non-dimensionalize the vortex strengths using the strength of the first vortex pair, $K_{1}$, and all distances by the initial average size of the three pairs, $y_{0}=(y_{1}^0+y_{2}^0+y_{3}^0)/3$. This yields six dimensionless parameters, as detailed in the caption of Fig.~\ref{fig:sensitivity}.

To assess the sensitivity of the period to the dimensionless parameters, we could sweep the entire six-dimensional parameter space and collect leapfrogging period data for all combinations. However, this approach is computationally intensive. Therefore, to address this, we develop a machine learning model to predict the period directly from the parameters, leveraging the power of neural networks.

Neural networks have shown great potential in modeling highly nonlinear and high-dimensional systems~\cite{colen2021machine, dulaney2021machine}. In this work, we design a fully connected feed-forward network composed of multiple layers. Each layer is of the form $f(Wx+b)$, where $f$ is a user-prescribed nonlinear activation function (chosen here as the sigmoid function), $x$ is the input vector of each layer, $W$ is the weight matrix and $b$ is the bias. Both $W$ and $b$ are learned during training. We also define a loss function: 
\begin{equation}\label{eq:loss}
    L=\sum\limits_{i=1}^{N}[(T^{i}_{12}-\widehat{T}^{i}_{12})^{2}+(T^{i}_{13}-\widehat{T}^{i}_{13})^{2}+(T^{i}_{23}-\widehat{T}^{i}_{23})^{2}],
\end{equation}
where $N$ is the total number of periodic leapfrogging cases that we train the neural network with, $T_{12}^i$ is the leapfrogging period for vortex pairs $1$--$2$ for case $i\in\{1,\cdots,N\}$, and $\widehat{T}^i_{12}$ is the corresponding value predicted from the neural network. The network is trained on the entire simulation dataset to update the weight matrix $W$ and bias vector $b$ by minimizing the loss function (equation~\eqref{eq:loss}) using the Adam optimizer. The datasets consist of 219 cases for vortex pairs 1--2, 196 for 1--3, and 135 for 2--3. 
After training, the neural network predicts $T_{\mathrm{leapfrog}}^{\mathrm{NN}}$ by varying one parameter while fixing the others. 
This allows us to identify which parameter has the largest effect on the period.

We present the results of our sensitivity analysis for the leapfrogging period in Fig.~\ref{fig:sensitivity}, where we use all six dimensionless parameters to plot the relative difference between the period predicted by the neural network ($T_{\mathrm{leapfrog}}^{\mathrm{NN}}$) and the period obtained by the simulations ($T_{\mathrm{leapfrog}}$), as a log-log plot, versus variations in the parameters.
Given the large number of parameters explored, the results are presented as a distribution of relative differences (thickness of each column) across all periodic cases, with thinner/wider parts representing fewer/more cases.

We observe significant increases in the relative differences when varying the parameters around their original values across the wide range of parameter variations studied (see Fig.~\ref{fig:sensitivity}). 
The black dots and yellow crosses within each violin plot represent the mean and median relative differences, respectively, and the orange lines connect the median values across all variations $\nu_i/\nu_i^{\star}$ for $i=1,2,\dots, 6$.
The violin plots illustrate the distribution of relative differences for each parameter variation.

To quantify the sensitivity, we compute the absolute slope of each orange line segment and calculate their average. Note that both axes are given on a logarithmic scale. 
The computed average slopes are approximately 0.23, 0.30, 0.33, 0.38, 1.21, and 0.98 for Figs.~\ref{fig:sensitivity}(a) through~(f), respectively. 
These average slopes are indicative of the sensitivity of $T_{\mathrm{leapfrog}}$ with respect to each parameter.
We find that the leapfrogging period is not as sensitive to the initial size of the vortex pairs (Figs.~\ref{fig:sensitivity}(a) and~(b)), it is more sensitive to the vortex strength (Figs.~\ref{fig:sensitivity}(c) and~(d)) and most sensitive to the initial horizontal distance between vortex pairs (Figs.~\ref{fig:sensitivity}(e) and~(f)). This is in good agreement with our earlier findings for two interacting coaxial vortex pairs, where we identified---numerically in Fig.~\ref{fig:twoRingsResults}(b) and analytically in Fig.~\ref{fig:Tleapfrog}(b)---the distance between the pairs as a dominant factor affecting the leapfrogging period. 
Additionally, we show that the vortex strength is another key factor influencing the period, which was previously unexamined since we set the strength to unity in the analysis of two interacting coaxial vortex pairs.

Interestingly, we observe that the relative differences generally exhibit uniform-like distributions (constant column thickness along the vertical axis), indicating that different parameter combinations exhibit varying sensitivity to the same parameter, despite following the same general trend. This suggests that while we clearly demonstrate the trend and quantify the sensitivity of the period to these parameters, further exploration is needed to better understand the complex relationships between them. Given the inherent nonlinearity of the period-parameter relationship, we plan to address this in future work.

\section{Summary and discussion}\label{sec:conclusions}

In this study we have analyzed interactions between two and three coaxial vortex pairs, classifying their dynamics as either ordered or chaotic based on strengths, initial sizes, and initial horizontal separations. Unlike previous works, we explored a broad range of initial geometric configurations and various strengths.

By examining the relative phase trajectories, we found that periodic cases are scattered among chaotic ones across different initial configurations. Quasi-periodic leapfrogging typically occurs when the initial distances between the vortex pairs are small and cannot coexist with vortex-pair overtake. When the initial configuration splits into two interacting vortex pairs and a single propagating vortex pair, the two interacting pairs consistently exhibit periodic leapfrogging.
For the smallest initial horizontal separations, the system predominantly exhibits chaotic or quasi-periodic motions rather than periodic leapfrogging with a single frequency. This behavior is due to the strong coupling between all three vortex pairs. When the pairs are in close proximity, more complex and chaotic dynamics emerge instead of periodic motion.

We quantified the occurrence of periodic leapfrogging by calculating its percentage out of all interacting cases, including quasi-periodic leapfrogging and chaotic interactions but excluding the overtake/pass-through cases, in the parameter space of vortex strength and initial size for the third pair. Our findings indicate that quasi-periodic leapfrogging and chaotic interactions generally occur when the three vortex pairs have similar strengths and initial sizes. Conversely, discrepancies in these parameters cause the system to disintegrate into two subsystems: a single propagating vortex pair and two periodically leapfrogging pairs.

Given the high-dimensional parameter space, we used a machine learning model using neural networks to perform a sensitivity analysis on the leapfrogging period, identifying initial horizontal separation as the most dominant factor. Future studies could further explore the stability and complex dynamics of these systems.

\vspace{.5cm}
 \begin{center}
   \textbf{Acknowledgements}  
\end{center}
C.M.\ thanks Alessandro Podo for useful discussions and acknowledges funding support provided by a Joseph~B.\ Keller Postdoctoral Fellowship at the Courant Institute at NYU. We thank Jiachen Huang for  discussions during the initial stages of this work.

\appendix

 \section{Derivation of the leapfrogging period for two vortex pairs of the same strength}\label{app:Tleapfrog}

We derive the leapfrogging period for two vortex pairs
following~\cite[App.\ B]{behring2020dances} and explicitly show how to use the derived formula for \emph{different initial horizontal separations} in addition to different initial sizes. By formulating the interacting point vortex equations given in equations~\eqref{eq:dxdt}--\eqref{eq:dydt} as a Hamiltonian system~\cite{aref2007point,aref2010150} we consider the special case of net-zero circulation for the 4-vortex problem, equivalent to two vortex rings of the same strength, whose Hamiltonian is given by equation~\eqref{eq:logHamiltonian}.

The Hamiltonian (equation~\eqref{eq:logHamiltonian}) becomes singular at $(X,Y)=(0,0)$ and to de-singularize it in this neighborhood we redefine the Hamiltonian using the transformation $\tilde{H}(\tilde{q}(\tau(t)),\tilde{p}(\tau(t))):=f(H(q(t),p(t)))$ where $f\in C^1(\mathbb{R})$ and is monotonic. This re-parametrizes time but the trajectories and level sets of the corresponding system coincide with the original Hamiltonian system.  From chain rule we can determine the new time parameter:
\begin{equation}
    \frac{\partial \tilde{H}}{\partial \tilde{q}}=f'(H)\frac{\partial H}{\partial q},
\end{equation}
and together with 
\begin{equation}
    \frac{\d q}{\d t}=\frac{\d \tilde{q}}{\d \tau}\frac{\d \tau}{\d t}\quad ;\quad \frac{\d p}{\d t}=\frac{\d \tilde{p}}{\d \tau}\frac{\d \tau}{\d t},
\end{equation}
we obtain from the equations of motion 
\begin{equation}
   \frac{\d q}{\d t}=\frac{\partial H}{\partial p}\quad ; \quad \frac{\d p}{\d t}=-\frac{\partial H}{\partial q},
\end{equation}
that 
\begin{equation}
   \frac{\d \tilde{q}}{\d \tau}\frac{\d \tau}{\d t}=-\frac{1}{f'(H)}\frac{\partial\tilde{H}}{\partial\tilde{p}} \quad ; \quad  \frac{\d \tilde{p}}{\d \tau}\frac{\d \tau}{\d t}=\frac{1}{f'(H)}\frac{\partial\tilde{H}}{\partial\tilde{q}}.
\end{equation}
Therefore, if the new time scale is described by 
\begin{equation}
    \frac{\d \tau}{\d t}=-\frac{1}{f'(H)},
\end{equation}
by integrating it with respect to $t$, we obtain
\begin{equation}
    \tau = -\frac{1}{f'(H)}t,
\end{equation}
and the Hamiltonian structure of the equations of motion is preserved. 

In our case, $f(H)$ will be $f(H)=\tilde{H}(X,Y)=e^{-2\pi H(X,Y)}/2$, which yields
\begin{equation}\label{eq:regHamiltonian}
    \tilde{H}(X,Y)=\frac{1}{2}\left(\frac{1}{1-Y^2}-\frac{1}{1+X^2}\right),
\end{equation}
and $\tau = e^{2\pi H}t/\pi$ with $H$ defined as in equation~\eqref{eq:logHamiltonian}.

\begin{figure}[htpb]
    \centering
    \includegraphics[width=.92\textwidth]{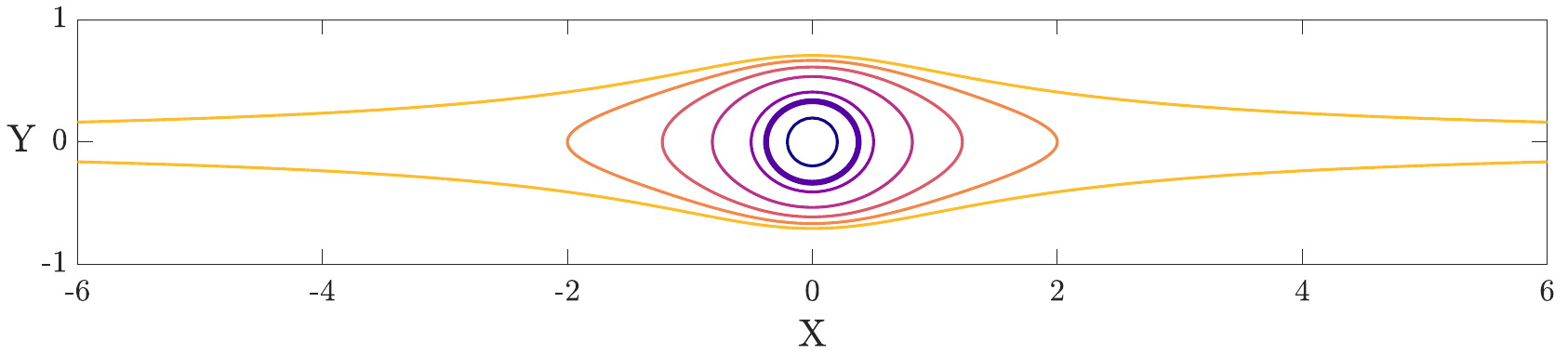}
    \caption{Level curves of the Hamiltonian (equation~\eqref{eq:regHamiltonian}) in the $X$-$Y$ plane, including the critical energy level $h =1/(2\tilde{H}_0)= 1/8$ (bold).}
    \label{fig:HamiltonianLevelCurves}
\end{figure}

The level curves of $H$ are shown in Fig.~\ref{fig:HamiltonianLevelCurves} (see also~\cite{love1893motion,tophoj2013instability}). We note here the symmetry about the $Y$-axis. In the particular case considered here, where $\mu=K_2/K_1=1$, the motions are symmetric about the $X$-axis as well. 

The new, modified Hamiltonian yields evolution equations given by
\begin{align}
    \frac{\d X}{\d \tau}&=\frac{\partial \tilde{H}}{\partial Y}=\frac{Y}{(1-Y^2)^2},\label{eq:simpledXdt}\\
    \frac{\d Y}{\d \tau}&=-\frac{\partial \tilde{H}}{\partial X}=-\frac{X}{(1+X^2)^2},
\end{align}
where for the $X$ and $Y$ here we have dropped the tildes for convenience.
This particular choice of initial condition corresponds to a specific value of $\tilde{H}_0=\tilde{H}(X_0,Y_0)$ and thus a particular level curve in Fig.~\ref{fig:HamiltonianLevelCurves}. Substituting $X=0$ and $Y=Y_0$ into equation~\eqref{eq:regHamiltonian} we get $2\tilde{H}_0=Y_0^2/(1-Y_0^2)$ and since we define
\begin{equation}\label{eq:hinH0}
    h=\frac{1}{2\tilde{H}_0},
\end{equation}
we can show that the last equality of equation~\eqref{eq:ICsHamiltonian} holds.
Since equation~\eqref{eq:hinH0} is related to the energy level and is conserved we have $\tilde{H}(X,Y)=\tilde{H}_0$.
Rearranging the initial condition $Y_0$ in equation~\eqref{eq:ICsHamiltonian} for $h$ we obtain its definition in terms of~$\alpha$: 
\begin{equation}\label{eq:hAndH0}
    h=\frac{4\alpha}{(1-\alpha)^2}\quad;\quad \tilde{H}_0=\frac{(1-\alpha)^2}{8\alpha}.
\end{equation}
We use equation~\eqref{eq:hinH0} in equation~\eqref{eq:regHamiltonian} to obtain $Y$ in terms of $X$ and $h$ as follows:
\begin{equation}
    Y^2=\frac{1}{h+1}\frac{1-(h-1)X^2}{1+\frac{X^2}{h+1}}.
\end{equation}
Taking this expression and plugging it into equation~\eqref{eq:simpledXdt} we get
\begin{equation}
    \frac{\d X}{\d \tau}=\frac{1}{h^2\sqrt{h+1}}\left(1-\frac{h}{h+1}\frac{1}{1+\frac{X^2}{h+1}}\right)^{-2}\sqrt{\frac{1-(h-1)X^2}{1+\frac{X^2}{h+1}}},
\end{equation}
which we can solve using separation of variables to obtain:
\begin{equation}
    \tau(X)=h^2\sqrt{h+1}\int_{0}^X\left(1-\frac{h}{h+1}\frac{1}{1+\frac{x^2}{h+1}}\right)^{2}\sqrt{\frac{1+\frac{x^2}{h+1}}{1-(h-1)x^2}}\,\d x.
\end{equation}
Next, we perform a change of variables: $u=x\sqrt{h-1}$, which yields
\begin{equation}\label{eq:tInU}
    \tau(X)=h^2\sqrt{\frac{h+1}{h-1}}\int_{0}^{X\sqrt{h-1}}\left(1-\frac{h}{h+1}\frac{1}{1-k^2u^2}\right)^{2}\sqrt{\frac{1-k^2u^2}{1-u^2}}\,\d u,
\end{equation}
where we use that $k^2=1/(1-h^2)$ and thus $u^2/(h^2-1)=-k^2u^2$.
We split the integrand of equation~\eqref{eq:tInU} into three separate parts:
\begin{align}
    &\left(1-\frac{h}{h+1}\frac{1}{1-k^2u^2}\right)^{2}\sqrt{\frac{1-k^2u^2}{1-u^2}}=\left[1-\frac{2h}{h+1}\frac{1}{1-k^2u^2}+\frac{h^2}{(h+1)^2}\frac{1}{(1-k^2u^2)^2}\right]\sqrt{\frac{1-k^2u^2}{1-u^2}}\\
    &=\sqrt{\frac{1-k^2u^2}{1-u^2}}-\frac{2h}{h+1}\frac{1}{\sqrt{1-k^2u^2}\sqrt{1-u^2}}+\frac{h^2}{(h+1)^2}\frac{1}{(1-k^2u^2)\sqrt{1-u^2}\sqrt{1-k^2u^2}}.
\end{align}
Using this decomposition into three fractions we now perform the integration 
\begin{align}
    \tau(X)=h^2\sqrt{\frac{h+1}{h-1}}&\left[\int_{0}^{X\sqrt{h-1}}\sqrt{\frac{1-k^2u^2}{1-u^2}}\,\d u-\frac{2h}{h+1}\int_{0}^{X\sqrt{h-1}}\frac{1}{\sqrt{1-k^2u^2}\sqrt{1-u^2}}\,\d u\right.\nonumber\\
    &\left.+\frac{h^2}{(h+1)^2}\int_{0}^{X\sqrt{h-1}}\frac{1}{(1-k^2u^2)}\frac{1}{\sqrt{1-k^2u^2}\sqrt{1-u^2}}\,\d u\right].
\end{align}
All these are incomplete elliptic integrals of the first, second, and third kind, so we can rewrite $\tau(X)$ as follows
\begin{equation}\label{eq:tXwithAllElliptic}
     \tau(X)=h^2\sqrt{\frac{h+1}{h-1}}\left[E(X\sqrt{h-1},k)-\frac{2h}{h+1}F(X\sqrt{h-1},k)+\frac{h^2}{(h+1)^2}\Pi (X\sqrt{h-1},k^2,k)\right],
\end{equation}
where we use $u=\sin(\theta)=x\sqrt{h-1}$. Our definition of $h$ in equation~\eqref{eq:hAndH0} implies that $h>1$.

We use the identity~\cite[19.6.13]{NIST:DLMF}: 
\begin{equation}
    \Pi(\varphi,k^2,k) = \frac{1}{1-k^2}\left( E(\varphi,k)-\frac{k^2\sin\varphi\cos\varphi}{\sqrt{1-k^2\sin^2\varphi}} \right),
\end{equation}
with $u=\sin\varphi = X\sqrt{h-1}$ (and thus, $\cos\varphi=\sqrt{1-\sin^2\varphi}=\sqrt{1-u^2}=\sqrt{1-X^2(h-1)}$) to write the incomplete elliptic integral of the third kind as a second kind one, through
\begin{equation}
    \Pi(x,k^2,k) = \frac{1}{1-k^2}\left( E(x,k)-\frac{k^2X\sqrt{h-1}\sqrt{1-X^2(h-1)}}{\sqrt{1-k^2X^2(h-1)}}\right).
\end{equation}
Using $k^2=1/(1-h^2)$ again, and upon a series of simplifications we obtain
\begin{equation}
    \Pi(x,k^2,k) = \frac{h^2-1}{h^2}E(x,k) + \sqrt{1-h^2}\frac{X\sqrt{X^2(h-1)-1}}{\sqrt{h+X^2+1}}.
\end{equation}
Plugging this expression into equation~\eqref{eq:tXwithAllElliptic} we remove the incomplete integral of the third kind and finally arrive at
\begin{equation}\label{eq:tauX}
    \tau(X) = \frac{2h^3}{\sqrt{h^2-1}}\left(E(X\sqrt{h-1},k) - F(X\sqrt{h-1},k) \right)+\frac{h^4}{h+1}\frac{X\sqrt{1-X^2(h-1)}}{\sqrt{h+X^2+1}}.
\end{equation}
Substituting $X=1/\sqrt{h-1}$ (which is equivalent to $\theta = \pi/2$), we get the quarter leapfrogging period in terms of the new time scale
\begin{equation}\label{eq:tau}
    \tau(X=1/\sqrt{h-1}) = \frac{2h^3}{\sqrt{h^2-1}}\left[E\left(\pi/2,k=\pm 1/\sqrt{1-h^2}\right) - F\left(\pi/2,k=\pm 1/\sqrt{1-h^2}\right)\right],
\end{equation}
where the last term in equation~\eqref{eq:tauX} vanishes when $X=1/\sqrt{h-1}$.
To obtain the leapfrogging period $T_{\mathrm{leapfrog}}$ in terms of the actual time scale we use $t=\pi e^{-2\pi H}\tau$ and multiply equation~\eqref{eq:tau} by 4:
\begin{equation}\label{eq:TleapfrogApp}
    T_{\mathrm{leapfrog}}(h) = 4(\pi e^{-2\pi H}\tau(X=1/\sqrt{h-1}))=\frac{\pi}{h} \left[\frac{8h^4}{h^2-1}E\left(\frac{1}{h} \right) - 8h^2K\left(\frac{1}{h}\right)\right],
\end{equation}
where $K$ and $E$ are the \textit{complete} elliptic integrals of the first and second kind, respectively. This is in agreement with Behring and Goodman~\cite{behring2019stability}, where the leapfrogging period is shown in the rescaled time variable. To obtain expressions with a real modulus we use the identities
\begin{equation}\label{eq:KandEidentities}
    K\left(\pm \frac{ik}{k'}\right) = k'K(k)\quad ; \quad E\left(\pm \frac{ik}{k'}\right) = \frac{1}{k'}E(k),
\end{equation}
where $k=1/h$ and $k'=\sqrt{h^2-1}/h$ so that $k^2+k'^2=1$. We note here that to compute $T_{\mathrm{leapfrog}}(h)$ in equation~\eqref{eq:TleapfrogApp} using the built-in \texttt{ellipke} function in \textsc{Matlab}, we have to use $1/h^2$ as the input of both $K$ and $E$ instead of $1/h$ (shown in equation~\eqref{eq:KandEidentities}),  because of the integral definition used in \texttt{ellipke}.

To compute analytically the leapfrogging period for arbitrary initial horizontal separations between the two vortex rings we make the following modifications. The initial condition $X(0)$ in equation~\eqref{eq:ICsHamiltonian} becomes 
\begin{equation}
    X(0)=X_0=\frac{x_1(0)-x_2(0)}{\hat{d}},
\end{equation}
which corresponds to a different level curve of $\tilde{H}_0$ (Fig.~\ref{fig:HamiltonianLevelCurves}) whose value is given by:
\begin{equation}
    \tilde{H}_0=\frac{1}{2}\left(\frac{1}{1-Y_0^2}-\frac{1}{1+X_0^2}\right).
\end{equation}


\section{Interactions of four coaxial vortex pairs}\label{app:fourPairs}

\begin{figure}[H]
    \centering
     \includegraphics[width=\textwidth]{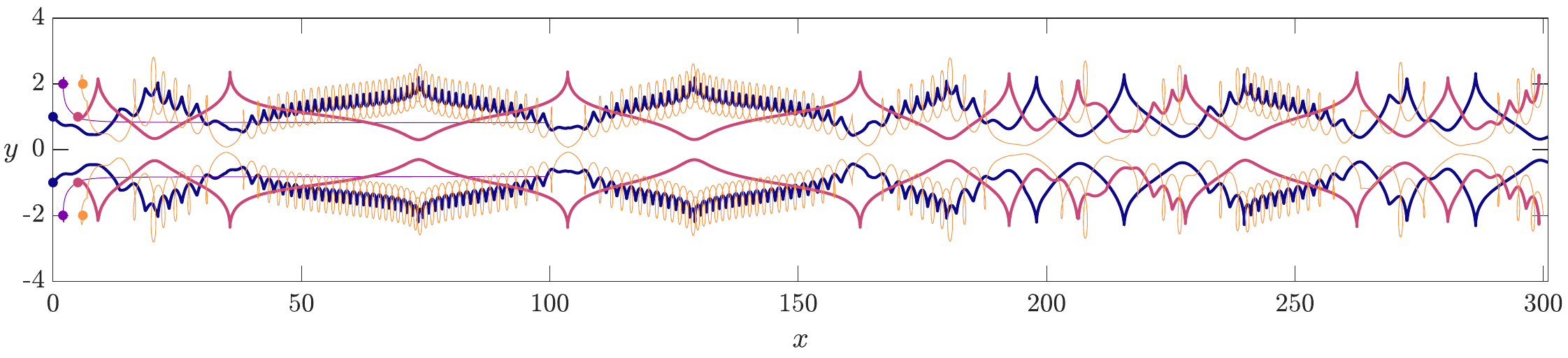}
     \caption{Example with four interacting vortex pairs (1: blue, 2: purple, 3: burgundy, 4: orange) with strengths $K_1=K_3=3$, $K_2=K_4=1$. The initial configuration is: $(x_1^0,y_1^0) = (0,1)$, $(x_2^0,y_2^0) = (2,2)$, $(x_3^0,y_3^0) = (5,1)$, and $(x_4^0,y_4^0) = (6,2)$.}
    \label{fig:example}
\end{figure}

The dynamics of systems comprising four or more coaxial vortex pairs can also be analyzed using the system of equations \eqref{eq:dxdt}--\eqref{eq:dydt}. Figure~\ref{fig:example} illustrates an example of four interacting vortex pairs, where the first and third vortex pairs have equal strengths, each three times that of the second and fourth vortex pairs. In this particular initial configuration of horizontal distances and sizes, the motion is chaotic, although periodic motions occur intermittently.

Future work could explore the potential for systems of more than three vortex pairs to exhibit quasi-periodic behavior. Konstantinov's 1997 study \cite{konstantinov1997numerical} showed that regular motion in systems of four or five vortex rings or pairs occurs only when they decompose into multiple subsystems. However, this conclusion was drawn from numerical simulations with a limited set of initial conditions and imposed symmetry, suggesting that some cases may have been overlooked.

 \bibliographystyle{elsarticle-num} \bibliography{biblio.bib}
\end{document}